\documentclass[10pt]{llncs}
\usepackage{graphicx}
\usepackage{url}

\usepackage{amsmath}
\usepackage{amssymb}

\def\Oh{\mbox{\rm O}}
\def\oh{\mbox{\rm o}}

\def\+{\!+\!}
\def\-{\!-\!}

\newcommand{\abs}[1]{\ensuremath{\lvert #1 \rvert}}
\newcommand{\avg}[1]{\ensuremath{\overline{ #1 }}}

\newcommand{\SA}{\ensuremath{\mathsf{SA}}}
\newcommand{\DA}{\ensuremath{\mathsf{DA}}}
\newcommand{\C}{\ensuremath{\mathsf{C}}}

\newcommand{\LCP}{\ensuremath{\mathsf{LCP}}}
\newcommand{\ILCP}{\ensuremath{\mathsf{ILCP}}}

\newcommand{\find}{\textsf{find}}
\newcommand{\locate}{\textsf{locate}}

\newcommand{\doclist}{\textsf{list}}
\newcommand{\topk}{\textsf{topk}}

\newcommand{\mrank}{\ensuremath{\mathsf{rank}}}

\newcommand{\mlcp}{\ensuremath{\mathsf{lcp}}}

\newcommand{\RePair}{\textsf{Re-Pair}}

\newcommand{\Brute}{\textsf{Brute}}
\newcommand{\BruteL}{\textsf{Brute-L}}
\newcommand{\BruteD}{\textsf{Brute-D}}
\newcommand{\Sada}{\textsf{Sada}}
\newcommand{\SadaCL}{\textsf{Sada-C-L}}
\newcommand{\SadaCD}{\textsf{Sada-C-D}}
\newcommand{\SadaIL}{\textsf{Sada-I-L}}
\newcommand{\SadaID}{\textsf{Sada-I-D}}
\newcommand{\WT}{\textsf{WT}}
\newcommand{\Grid}{\textsf{Grid}}
\newcommand{\Grammar}{\textsf{Grammar}}
\newcommand{\LZ}{\textsf{LZ}}
\newcommand{\PDL}{\textsf{PDL}}
\newcommand{\PDLBC}{\textsf{PDL-BC}}
\newcommand{\PDLRP}{\textsf{PDL-RP}}
\newcommand{\PDLset}{\textsf{PDL-set}}
\newcommand{\PDLtopk}[1]{\textsf{PDL-#1}}

\newcommand{\Enwiki}{\textsf{Enwiki}}
\newcommand{\Page}{\textsf{Page}}
\newcommand{\Revision}{\textsf{Revision}}
\newcommand{\Influenza}{\textsf{Influenza}}
\newcommand{\Swissprot}{\textsf{Swissprot}}
\newcommand{\DNA}{\textsf{DNA}}
\newcommand{\Concat}{\textsf{Concat}}
\newcommand{\Version}{\textsf{Version}}

\title{Document Retrieval on Repetitive Collections\thanks{This work is funded in part by: 
Fondecyt Project 1-140796 (first author); Basal Funds FB0001, Conicyt, Chile 
(first and third authors); the Jenny and Antti Wihuri Foundation, Finland (third author); and 
by the Academy of Finland through grants 258308 and 250345 (CoECGR) (second author).}}

\author{
Gonzalo Navarro\inst{1} 
\and
Simon J. Puglisi\inst{2}
\and
Jouni Sir\'en\inst{1} 
}

\institute{
    Center for Biotechnology and Bioengineering, Department of Computer Science,
    University of Chile, Chile\\
    \email{\{gnavarro,jsiren\}@dcc.uchile.cl}\\[1ex]
\and
    Department of Computer Science,
    University of Helsinki, Finland\\
    \email{puglisi@cs.helsinki.fi}\\[1ex]
}

\date{}

\begin{document}

\maketitle

\begin{abstract}
Document retrieval aims at finding the most important documents
where a pattern appears in a collection of strings. Traditional
pattern-matching techniques yield brute-force document retrieval solutions,
which has motivated the research on tailored indexes that offer near-optimal 
performance. However, an experimental study establishing which alternatives
are actually better than brute force, and which perform best depending on the
collection characteristics, has not been carried out.
In this paper we address this shortcoming by exploring the relationship 
between the nature of the underlying collection and the performance of current 
methods. Via extensive experiments we show that established 
solutions are often beaten in practice by brute-force alternatives. We also
design new methods that offer superior time/space trade-offs, particularly on
repetitive collections.
\end{abstract}

\section{Introduction}

The {\em pattern matching} problem, that is, preprocessing a text collection 
so as to efficiently find the occurrences of patterns, is a classic in Computer 
Science. The optimal suffix tree solution~\cite{Wei73} dates back to 1973. 
Suffix arrays \cite{MM93} are a simpler, near-optimal alternative.
Surprisingly, the natural variant of the problem called
{\em document listing}, where one wants to find simply in which texts of the 
collection (called the {\em documents}) a pattern appears, was not solved
optimally until almost 30 years later \cite{Mut02}. Another natural variant, 
the {\em top-$k$ documents} problem, where one wants to find the $k$ {\em
most relevant} documents where a pattern appears, for some notion of
relevance, had to wait for other 10 years \cite{HSV09,NN12}.


A general problem with the above indexes is their size. While for moderate-sized collections (of total length $n$)
their linear space (i.e., $\Oh(n)$ words, or $\Oh(n\log n)$
bits) is affordable, the constant factors multiplying the linear term make the
solutions prohibitive on large collections. In this aspect, again, the pattern
matching problem has had some years of advantage. The first compressed suffix
arrays (CSAs) appeared in the year 2000 (see~\cite{NM07}) and since then have
evolved until achieving, for example, asymptotically optimal space 
in terms of high-order empirical entropy and time slightly over the optimal.
There has been much research on similarly compressed data structures for document 
retrieval (see \cite{NavACMcs14}). Since the foundational paper of Hon et 
al.~\cite{HSV09}, results have come close to using just $\oh(n)$ bits on top of 
the space of a CSA and almost optimal time.

Compressing in terms of statistical entropy is adequate in many cases, but it
fails in various types of modern collections.
{\em Repetitive}
document collections, where most documents are similar, in whole or piecewise,
to other documents, naturally arise in fields like computational biology, 
versioned collections, periodic publications, and software repositories (see~\cite{Naviwoca12}). 
The
successful pattern matching indices for these types of collections use
grammar or Lempel-Ziv compression, which exploit repetitiveness
\cite{CN12,FN13}.
There are only a couple of document listing indices for repetitive
collections \cite{GKNPS13,CM13}, and
none for the top-$k$ problem.

Although several document retrieval solutions have been implemented and tested 
in practice \cite{NV12,KN13,FN13,GKNPS13}, no systematic practical study of
how these indexes perform, depending on the collection characteristics, has
been carried out. 

A first issue is to determine under what circumstances specific document listing solutions actually beat
brute-force solutions based on pattern matching.
In many applications documents are relatively small (a few kilobytes)
and therefore are unlikely to contain many occurrences of a given pattern.
This means that in practice the number of pattern occurrences ($occ$) may not be much
larger than the number of documents the pattern occurs in ($docc$), and therefore pattern matching-based solutions may be
competitive.

A second issue that has been generally neglected in the literature is that collections have
different kinds of repetitiveness, depending on the
application. For example, one might have a set of distinct documents, each one
internally repetitive piecewise, or a set of documents that are in whole
similar to each other. The repetition structure can be linear (each document
similar to a previous one) as in versioned collections, or even tree-like, or
completely unstructured, as in some biological collections. It is not clear
how current document retrieval solutions behave depending on the type of 
repetitiveness.

In this paper we carry out a thorough experimental study of the performance
of most existing solutions to document listing and top-$k$
document retrieval, considering various types of real-life and synthetic 
collections. We show that brute-force solutions are indeed competitive in
several practical scenarios, and that some existing solutions perform well
only on some kinds of repetitive collections, whereas others present a more
stable behavior. We also design new and superior alternatives for top-$k$ document
retrieval.

\section{Background}\label{section:background}

Let $T[1,n]$ be a concatenation of a collection of $d$ documents. We assume each document ends with a special character $\$$ that is lexicographically smaller than any other character of the alphabet. The \emph{suffix array} of the collection is an array $\SA[1,n]$ of pointers to the suffixes of $T$ in lexicographic order. The \emph{document array} $\DA[1,n]$ is a related array, where $\DA[i]$ is the identifier of the document containing $T[\SA[i]]$. Let $B[1,n]$ be a bitvector, where $B[i]=1$ if a new document begins at $T[i]$. We can map text positions to document identifiers by: $\DA[i] = \mrank_{1}(B,\SA[i])$, where $\mrank_{1}(B,j)$ is the number of $1$-bits in prefix $B[1,j]$.

In this paper, we consider indexes supporting four kinds of queries: 1) \find($P$) returns the range $[sp,ep]$, where the suffixes in $\SA[sp,ep]$ start with pattern $P$; 2) \locate($sp,ep$) returns $\SA[sp,ep]$; 3) \doclist($P$) returns the identifiers of documents containing pattern $P$; and 4) \topk($P,k$) returns the identifiers of the $k$ documents containing the most occurrences of $P$. CSAs support the first two queries. \find() is relatively fast, while \locate() can be much slower. The main time/space trade-off in a CSA, the \emph{suffix array sample period}, affects the performance of \locate() queries. 
Larger sample periods result in slower and smaller indexes. 

Muthukrishnan's document listing algorithm~\cite{Mut02} uses an array
$\C[1,n]$, where $\C[i]$ points to the last occurrence of $\DA[i]$ in
$\DA[1,i-1]$. Given a query range $[sp,ep]$, $\DA[i]$ is the first occurrence
of that document in the range iff $\C[i] < sp$. A \emph{range minimum query}
(RMQ) structure over $\C$ is used to find the position $i$ with the smallest
value in $\C[sp,ep]$. If $\C[i] < sp$, the algorithm reports $\DA[i]$, and
continues recursively in $[sp,i-1]$ and $[i+1,ep]$. Sadakane~\cite{Sad07}
improved the space usage with two observations: 1) if the recursion is done in
preorder from left to right, $\C[i] \ge sp$ iff document $\DA[i]$ has been
seen before, so array $\C$ is not needed; and 2) array $\DA$ can also be removed by using \locate() and $B$ instead.

Let $\mlcp(S,T)$ be the length of the \emph{longest common prefix} of
sequences $S$ and $T$. The LCP array of $T[1,n]$ is an array $\LCP[1,n]$,
where $\LCP[i] = \mlcp(T[\SA[i-1],n], T[\SA[i],n])$. We obtain the
\emph{interleaved LCP array} $\ILCP[1,n]$ by building separate LCP arrays for
each of the documents, and interleaving them according to the document array.
As $\ILCP[i] < \abs{P}$ iff position $i$ contains the first occurrence of
$\DA[i]$ in $\DA[sp,ep]$, we can use Sadakane's algorithm with RMQs over $\ILCP$ instead of $\C$~\cite{GKNPS13}. If the collection is repetitive, we can get a smaller and faster index by building the RMQ only over the run heads in $\ILCP$.

\section{Algorithms}\label{section:algorithms}

In this section we review {\em practical} methods for document listing and top-$k$ document retrieval.
For a more detailed review see, e.g., \cite{NavACMcs14}.



\noindent {\bf Brute force.} These algorithms sort the document identifiers in range $\DA[sp,ep]$ and report each of them once. \BruteD{} stores $\DA$ in $n \log d$ bits, while \BruteL{} retrieves the range $\SA[sp,ep]$ with \locate()  and uses bitvector $B$ to convert it to $\DA[sp,ep]$. Both algorithms can also be used for top-$k$ retrieval by computing the frequency of each document identifier and then sorting by frequency.

\noindent {\bf Sadakane.}
This is a family of algorithms based on Sadakane's improvements~\cite{Sad07} to Muthukrishnan's algorithm~\cite{Mut02}. \SadaCL{} is the original algorithm of Sadakane, while \SadaCD{} uses an explicit document array instead of retrieving the document identifiers with \locate(). \SadaIL{} and \SadaID{} are otherwise the same, respectively, except that they build the RMQ over $\ILCP$~\cite{GKNPS13} instead of $\C$.

\noindent {\bf Wavelet tree.} A \emph{wavelet tree} over a sequence can be used to quickly list the distinct values in any substring, and hence a wavelet tree over $\DA$ can be a good solution for many document retrieval problems. The best known implementation of wavelet tree-based document listing~\cite{NV12} can use plain, entropy-compressed~\cite{NM07}, and grammar-compressed~\cite{LM00} bitvectors in the wavelet tree. Our \WT{} uses a heuristic similar to the original WT-alpha~\cite{NV12}, multiplying the size of the plain bitvector by $0.81$ and the size of the entropy-compressed bitvector by $0.9$, before choosing the smallest one for each level of the tree.

For top-$k$ retrieval, \WT{} combines the wavelet tree used in document listing with a space-efficient implementation~\cite{NV12} of the top-$k$ trees of Hon et al.~\cite{HSV09}. Out of the alternatives investigated by Navarro and Valenzuela~\cite{NV12}, we tested the greedy algorithm, LIGHT and XLIGHT encodings for the trees, and sampling 
parameter $g' = 400$. 
In the results, we use the slightly smaller XLIGHT.

\noindent {\bf Precomputed document listing.} \PDL~\cite{GKNPS13} builds a
sparse suffix tree for the collection, and stores the answers to document
listing queries for the nodes of the tree. For long query ranges, we compute
the answer to the \doclist() query as a union of a small number of stored
answer sets. The answers for short ranges are computed by using \BruteL. \PDLBC{} is the original version, using a web graph compressor~\cite{HNspire12.3} to compress the sets. If a subset $S'$ of document identifiers occurs in many of the stored sets, the compressor creates a grammar rule $X \to S'$, and replaces the subset with $X$. We chose block size $b=256$ and storing factor $\beta=16$ as good general-purpose parameter values. 
We extend \PDL{} in Section~\ref{section:pdl}.

\noindent {\bf Grammar-based.} \Grammar~\cite{CM13} is an adaptation of a grammar-compressed self-index~\cite{CN12} for document listing. Conceptually similar to \PDL, \Grammar{} uses \RePair~\cite{LM00} to parse the collection. For each nonterminal symbol in the grammar, it stores the set of document identifiers whose encoding contains the symbol. A second round of \RePair{} is used to compress the sets. Unlike most of the other solutions, \Grammar{} is an independent index and needs no CSA to operate.

\noindent {\bf Lempel-Ziv.}
\LZ~\cite{FN13} is an adaptation of self-indexes based on LZ78 parsing for document listing. Like \Grammar, \LZ{} does not need a CSA.


\noindent {\bf Grid.} \Grid~\cite{KN13} is a faster but usually larger alternative to \WT. It can answer top-$k$ queries quickly if the pattern occurs at least twice in each reported document. If documents with just one occurrence are needed, \Grid{} uses a variant of \SadaCL{} to find them. We also tried to use \Grid{} for document listing, but the performance was not good, as it usually reverted to \SadaCL.


\section{Extending Precomputed Document Listing}\label{section:pdl}

In addition to \PDLBC, we implemented another variant of precomputed document listing~\cite{GKNPS13} that uses \RePair{}~\cite{LM00} instead of the biclique-based compressor. 

In the new variant, named \PDLRP, each stored set is represented as an increasing sequence of document identifiers. The stored sets are compressed with \RePair, but otherwise \PDLRP{} is the same as \PDLBC.
Due to the multi-level grammar generated by \RePair, decompressing the sets can be slower in \PDLRP{} than in \PDLBC. Another drawback comes from representing the sets as sequences: when the collection is non-repetitive, \RePair{} cannot compress the sets very well. 
On the positive side, compression is much faster and more stable. 

We also tried an intermediate variant, \PDLset, that uses \RePair-like set compression. While ordinary \RePair{} replaces common substrings $ab$ of length $2$ with grammar rules $X \to ab$, the compressor used in \PDLset{} searches for symbols $a$ and $b$ that occur often in the same sets. 
Treating the sets this way should lead to better compression on non-repetitive collections, but 
unfortunately our current compression algorithm is still too slow with non-repetitive collections.
With repetitive collections, the size of \PDLset{} is very similar to \PDLRP.

Representing the sets as sequences allows for storing the document identifiers
in any desired order. One interesting order is the top-$k$ order: store the
identifiers in the order they should be returned by a \topk() query. This
forms the basis of our new \PDL{} structure for top-$k$ document retrieval. In
each set, document identifiers are sorted by their frequencies in decreasing
order, with ties broken by sorting the identifiers in increasing order. The
sequences are then compressed by \RePair. If document frequencies are needed,
they are stored in the same order as the identifiers. The frequencies can be represented space-efficiently by first run-length encoding the sequences, and then using differential encoding for the run heads. If there are $b$ suffixes in the subtree corresponding to the set, there are $\Oh(\sqrt{b})$ runs, so the frequencies can be encoded in $\Oh(\sqrt{b} \log b)$ bits.

There are two basic approaches to using the \PDL{} structure for top-$k$ document retrieval. We can set $\beta = 0$, storing the document sets for all suffix tree nodes above the leaf blocks. This approach is very fast, as we need only decompress the first $k$ document identifiers from the stored sequence. It works well with repetitive collections, while the total size of the document sets becomes too large with non-repetitive collections. We tried this approach with block sizes $b = 64$ (\PDLtopk{64} without frequencies and \PDLtopk{64+F} with frequencies) and $b = 256$ (\PDLtopk{256} and \PDLtopk{256+F}).

Alternatively, we can build the \PDL{} structure normally with $\beta > 1$, achieving better compression. Answering queries is now slower, as we have to decompress multiple document sets with frequencies, merge the sets, and determine the 
top $k$.
We tried different heuristics for merging only prefixes of the document sequences, stopping when a correct answer to the top-$k$ query could be guaranteed. The heuristics did not generally work well, making brute-force merging the fastest alternative. We used block size $b = 256$ and storing factors $\beta = 2$ (\PDLtopk{256-2}) and $\beta = 4$ (\PDLtopk{256-4}). Smaller block sizes increased both index size and query times, as the number of sets to be merged was generally larger.

\section{Experimental Data}\label{section:data}

We did extensive experiments with both real and synthetic
collections.\footnote{See \url{http://www.cs.helsinki.fi/group/suds/rlcsa/}
for datasets and full results.}
The details of the collections can be seen in Table~\ref{table:collections} in the Appendix,
where we also describe how the search patterns were obtained.

Most of our document collections were relatively small, around 100~MB in size, as the \WT{} implementation uses 32-bit libraries, while \Grid{} requires large amounts of memory for index construction. We also used larger versions of some collections, up to 1~GB in size, to see how collection size affects the results. In practice, collection size was more important in top-$k$ document retrieval, as increasing the number of documents generally increases the $docc/k$ ratio. In document listing, document size is more important than collection size, as the performance of \Brute{} depends on the $occ/docc$ ratio.

\noindent {\bf Real collections.}
\Page{} and \Revision{} are repetitive collections generated from a Finnish language Wikipedia archive with full version history. The collection consists of either $60$ pages (small) or $280$ pages (large), with a total of $8834$ or $65565$ revisions. In $\Page$, all revisions of a page form a single document, while each revision becomes a separate document in $\Revision$.
\Enwiki{} is a nonrepetitive collection of $7000$ or $90000$ pages from a snapshot of the English language Wikipedia. 
\Influenza{} is a repetitive collection containing the genomes of $100000$ or $227356$ influenza viruses. 
\Swissprot{} is a nonrepetitive collection of $143244$ protein sequences used
in many document retrieval papers~(e.g.,~\cite{NV12}). As the full collection is only 54~MB, there is no large version of \Swissprot.

\noindent {\bf Synthetic collections.}
To explore the effect of 
collection repetitiveness on document retrieval performance in more detail, we generated three types of
synthetic collections, using files from the Pizza \& Chilli corpus%
\footnote{\url{http://pizzachili.dcc.uchile.cl/}}.

\DNA{} is similar to \Influenza. Each collection has $1$, $10$, $100$, or $1000$ base documents, $100000$, $10000$, $1000$, or $100$ variants of each base document, respectively, and mutation rate $p = 0.001$, $0.003$, $0.01$, $0.03$, or $0.1$. We generated the base documents by mutating a sequence of length $1000$ from the DNA file with zero-order entropy preserving point mutations, with probability $10p$. We then generated the variants in the same way with mutation rate $p$.

\Concat{} is similar to \Page. We read $10$, $100$, or $1000$ base documents of length $10000$ from the English file, and generated $1000$, $100$, or $10$ variants of each base document, respectively. The variants were generated by applying zero-order entropy preserving point mutations with probability $0.001$, $0.003$, $0.01$, $0.03$, or $0.1$ to the base document, and all variants of a base document were concatenated to form a single document. We also generated collections similar to \Revision{} by making each variant a separate document. These collections are called \Version{}.

\section{Experimental Results}\label{section:experiments}


We implemented \Brute, \Sada, and \PDL{} ourselves\footnote{Available at \url{http://www.cs.helsinki.fi/group/suds/rlcsa/}}, and modified existing implementations of \WT, \Grid, \Grammar, and \LZ{} for our purposes. All implementations were written in C++. Details of our test machine 
are in the Appendix.

As our CSA, we used RLCSA~\cite{Maekinen2010}, a practical implementation of a CSA that compresses repetitive collections well. The \locate() support in RLCSA includes optimizations for long query ranges and repetitive collections, which is important for \BruteL{} and \SadaIL. We used suffix array sample periods $8, 16, 32, 64, 128$ for non-repetitive collections and $32, 64, 128, 256, 512$ for repetitive ones.

For algorithms using a CSA, we broke the \doclist($P$) and \topk($P,k$) queries into a \find($P$) query, followed by a \doclist($[sp,ep]$) query or \topk($[sp,ep],k$) query, respectively. The measured times do not include the time used by the \find() query. As this time is common to all solutions using a CSA, and negligible compared to the time used by \Grammar{} and \LZ, the omission does not affect the results.

\begin{figure}[p]
\minipage{0.49\textwidth}
  \includegraphics[trim = 0mm 25mm 0mm 0mm, width=\linewidth]{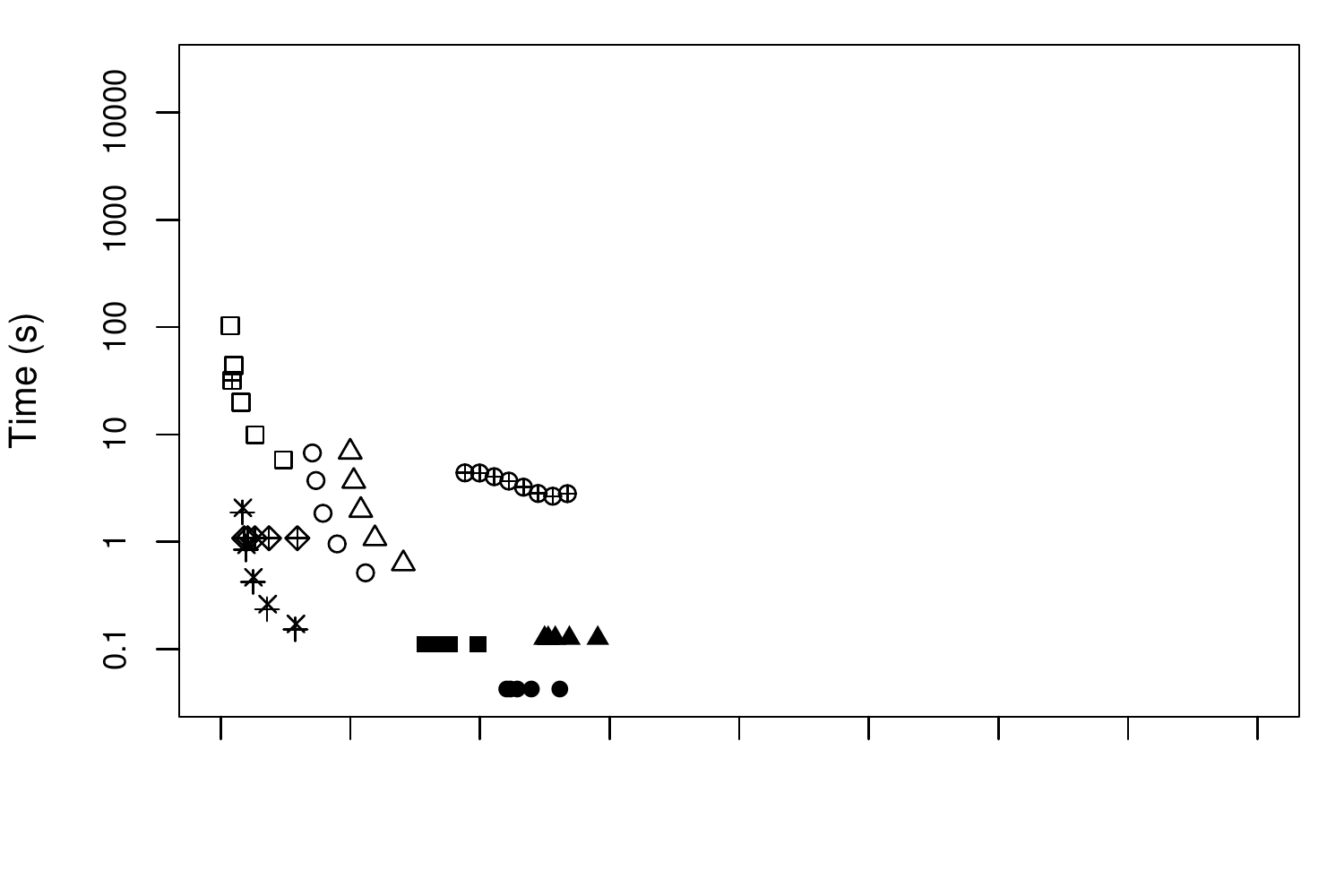}
\endminipage\hfill
\minipage{0.49\textwidth}
  \includegraphics[trim = 20mm 25mm -20mm 0mm, width=\linewidth]{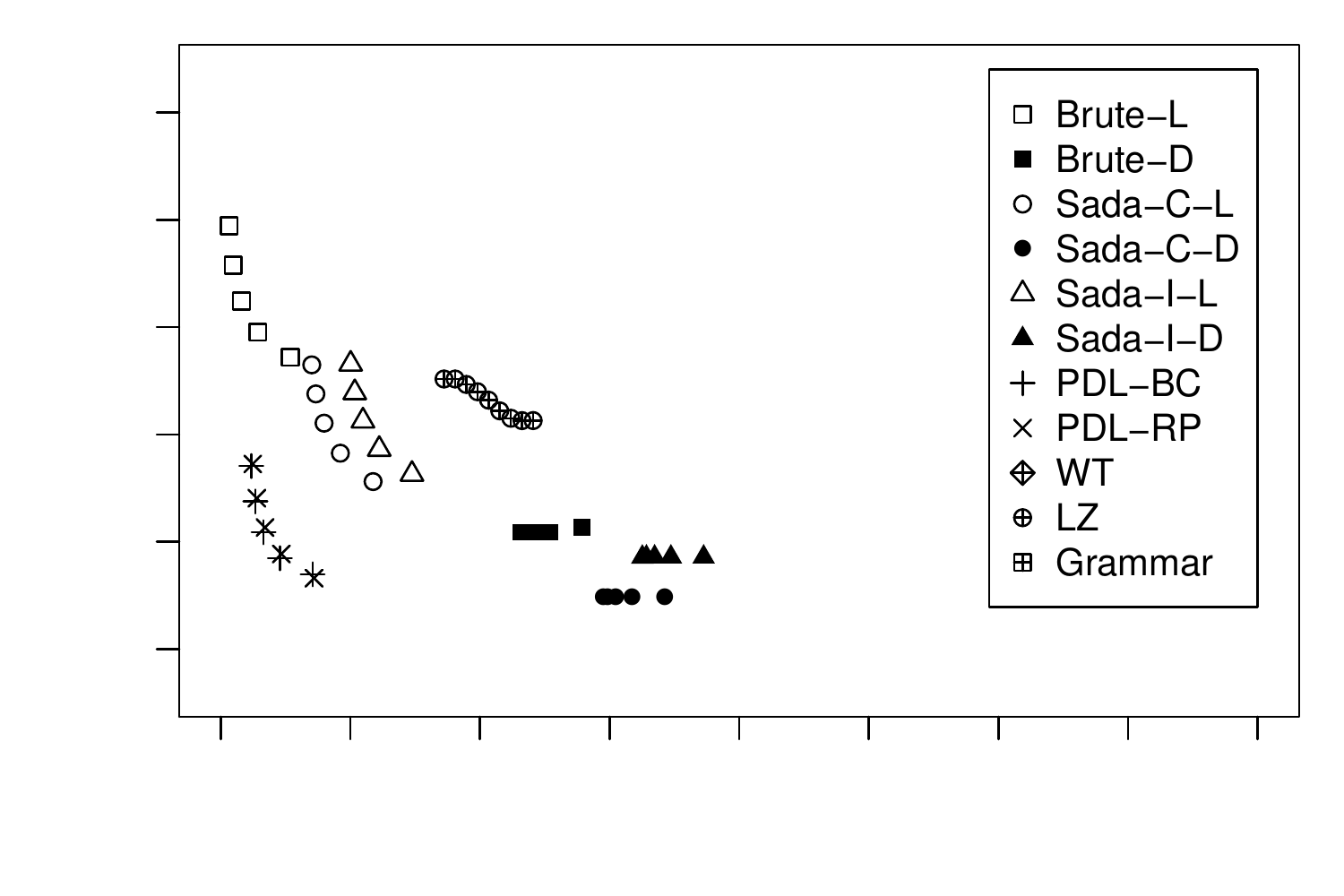}
\endminipage
\vspace{1ex}
\newline
\minipage{0.49\textwidth}
  \includegraphics[trim = 0mm 25mm 0mm 0mm, width=\linewidth]{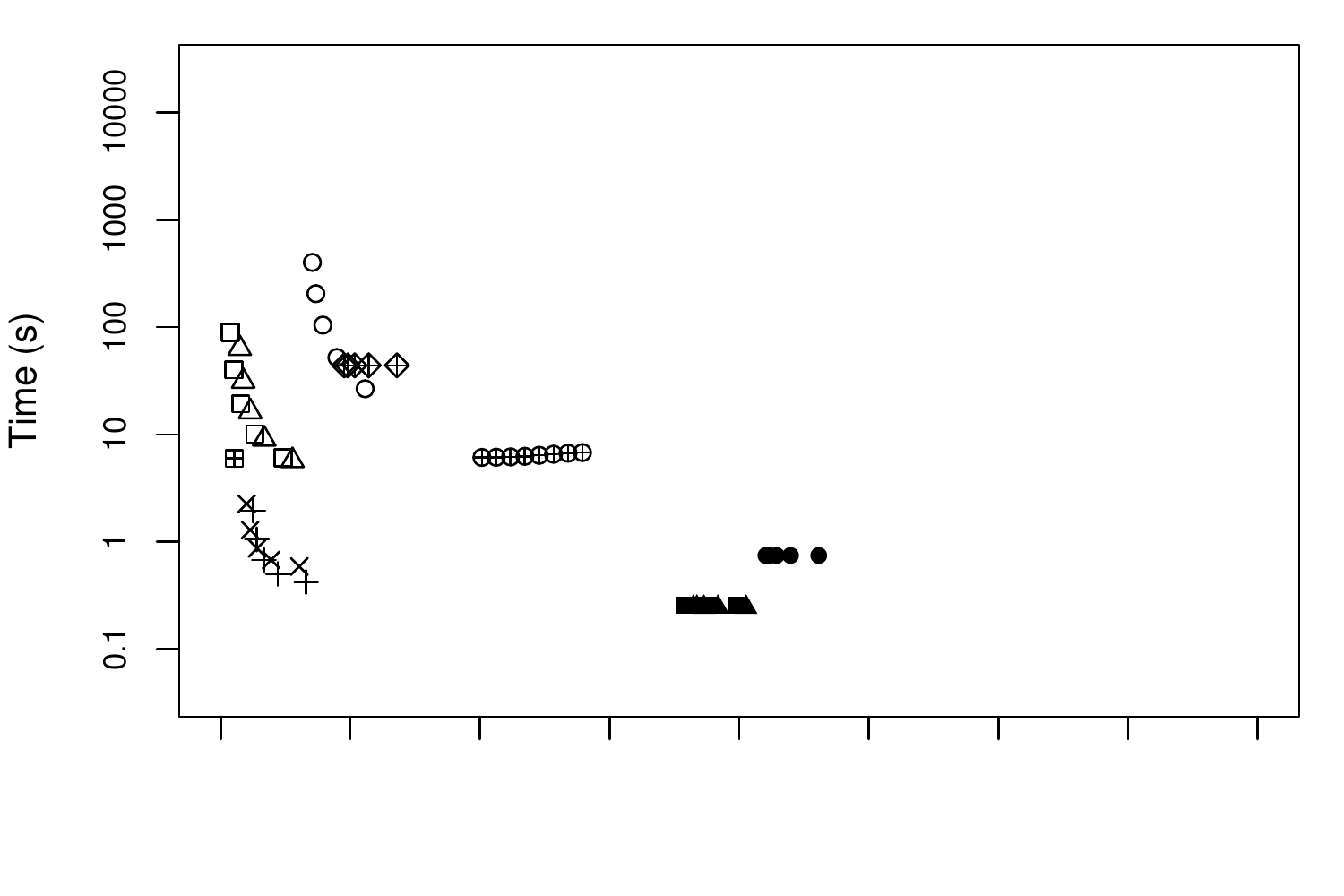}
\endminipage\hfill
\minipage{0.49\textwidth}
  \includegraphics[trim = 20mm 25mm -20mm 0mm, width=\linewidth]{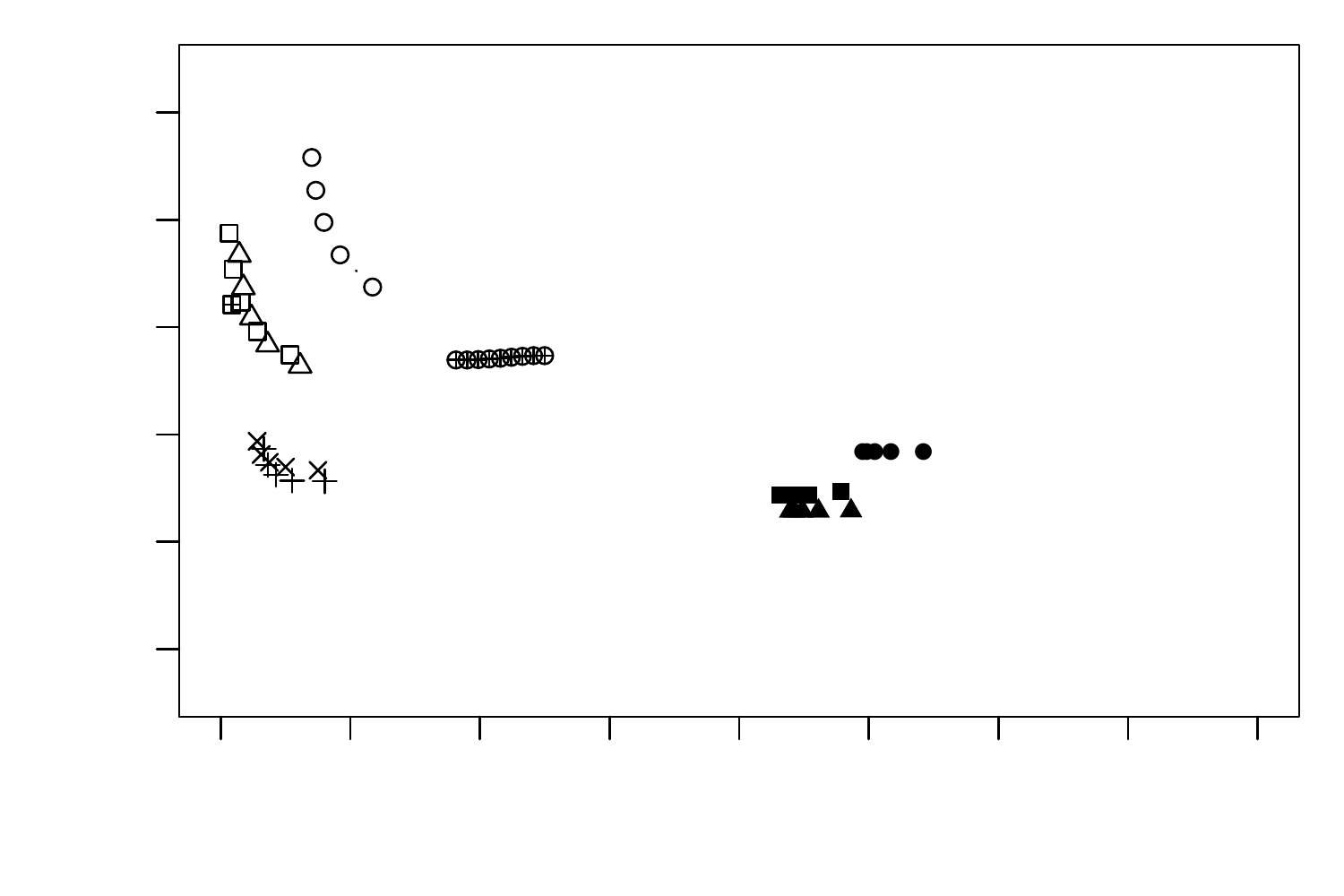}
\endminipage
\vspace{1ex}
\newline
\minipage{0.49\textwidth}
  \includegraphics[trim = 0mm 25mm 0mm 0mm, width=\linewidth]{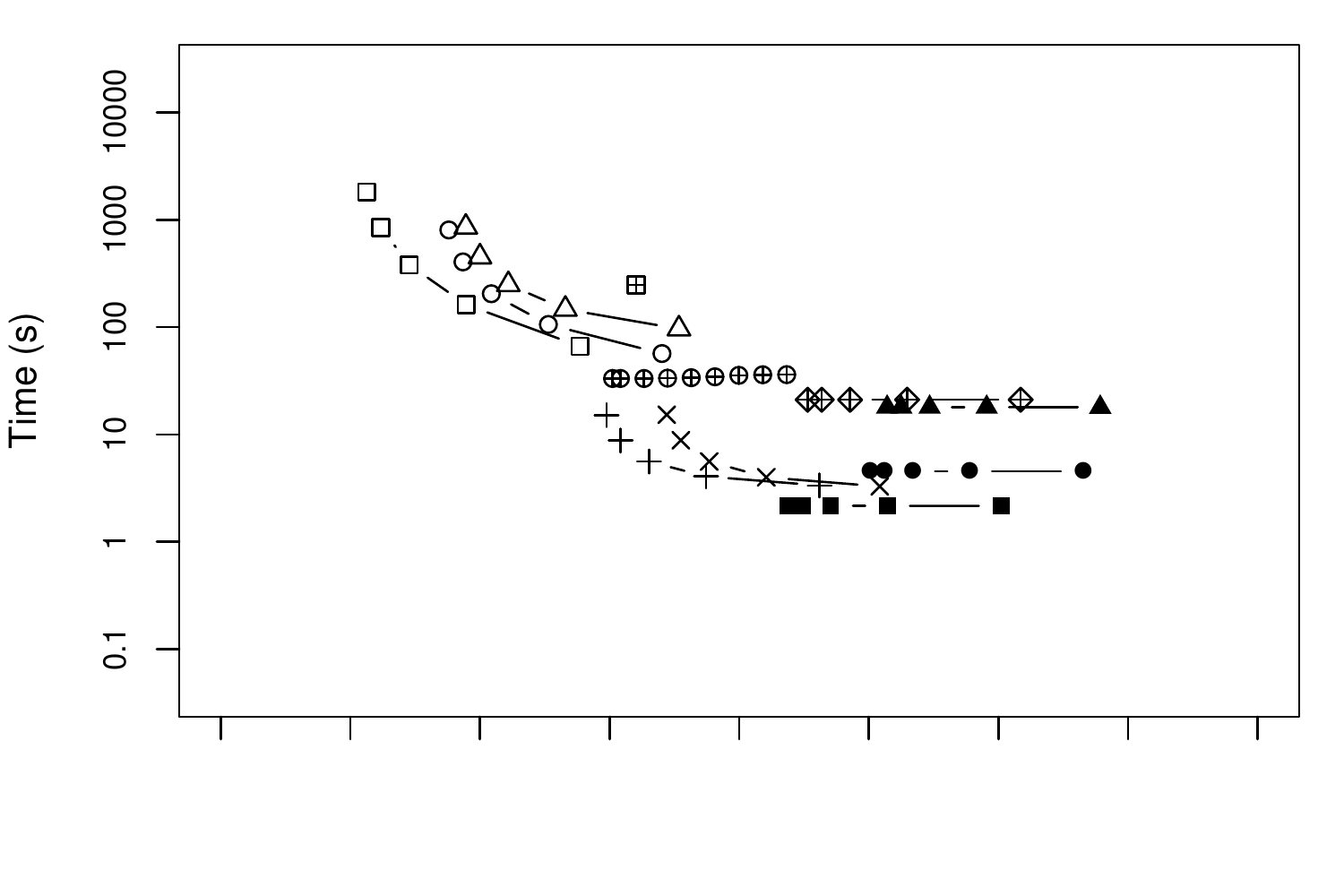}
\endminipage\hfill
\minipage{0.49\textwidth}
  \includegraphics[trim = 20mm 25mm -20mm 0mm, width=\linewidth]{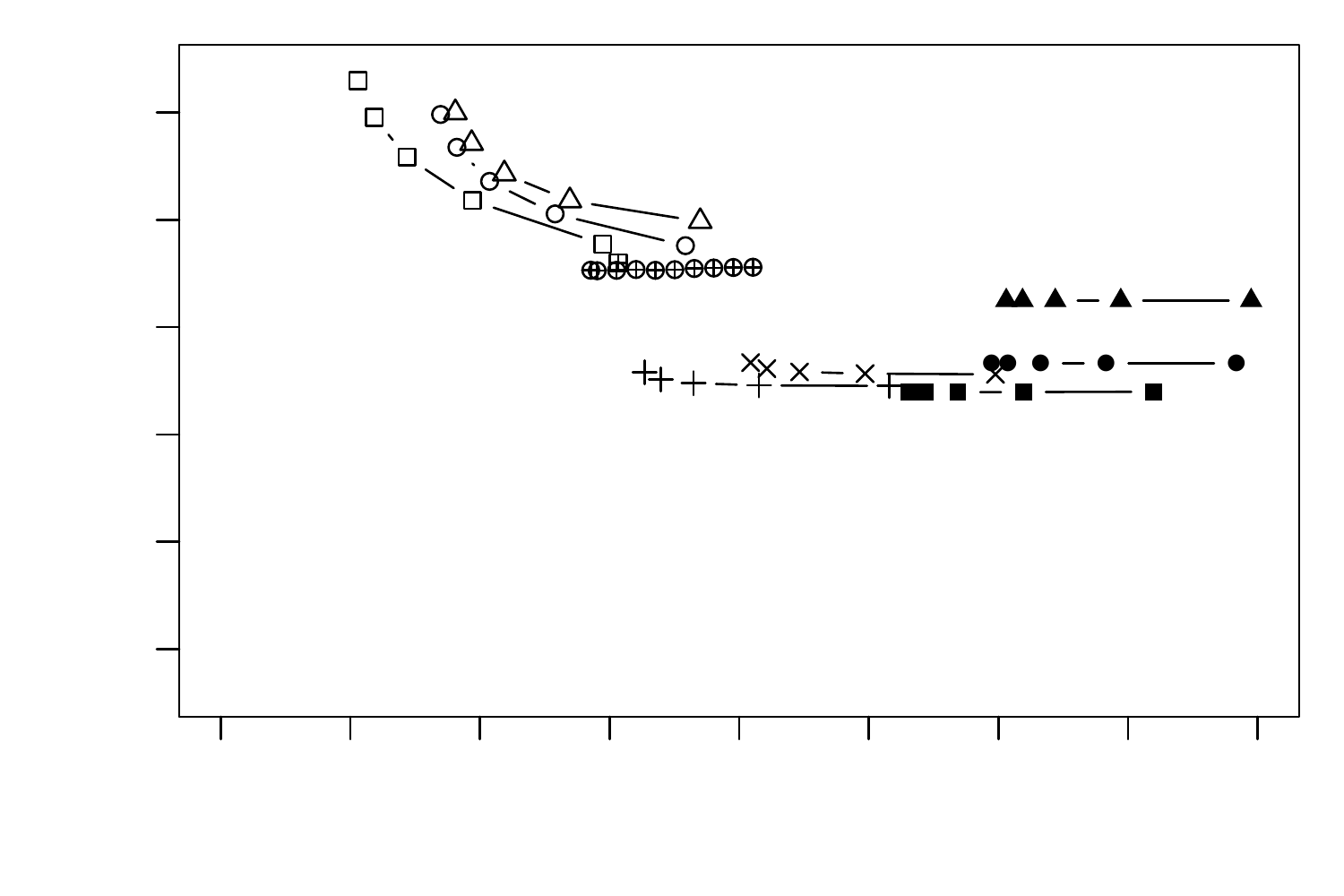}
\endminipage
\vspace{1ex}
\newline
\minipage{0.49\textwidth}
  \includegraphics[trim = 0mm 25mm 0mm 0mm, width=\linewidth]{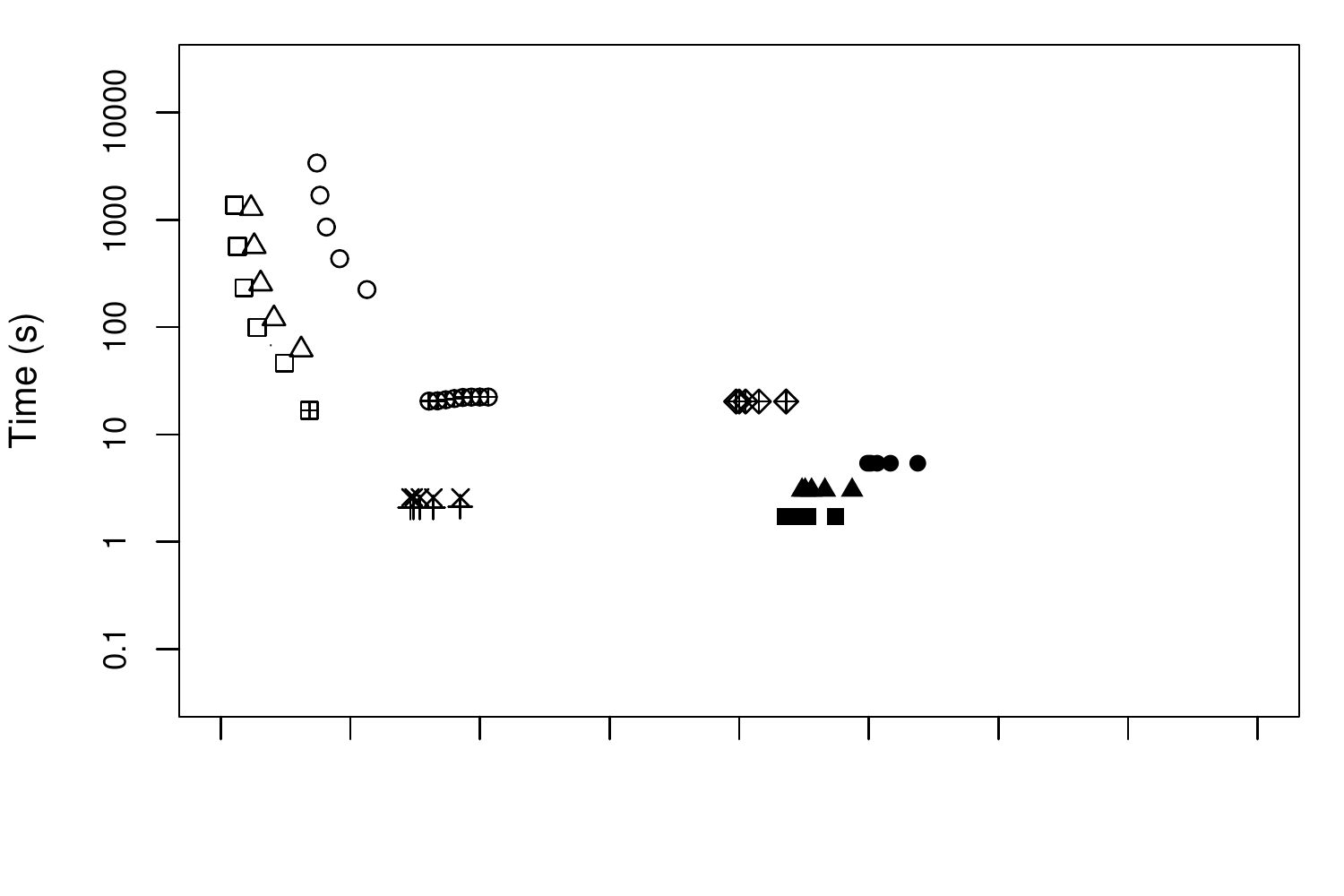}
\endminipage\hfill
\minipage{0.49\textwidth}
  \includegraphics[trim = 20mm 25mm -20mm 0mm, width=\linewidth]{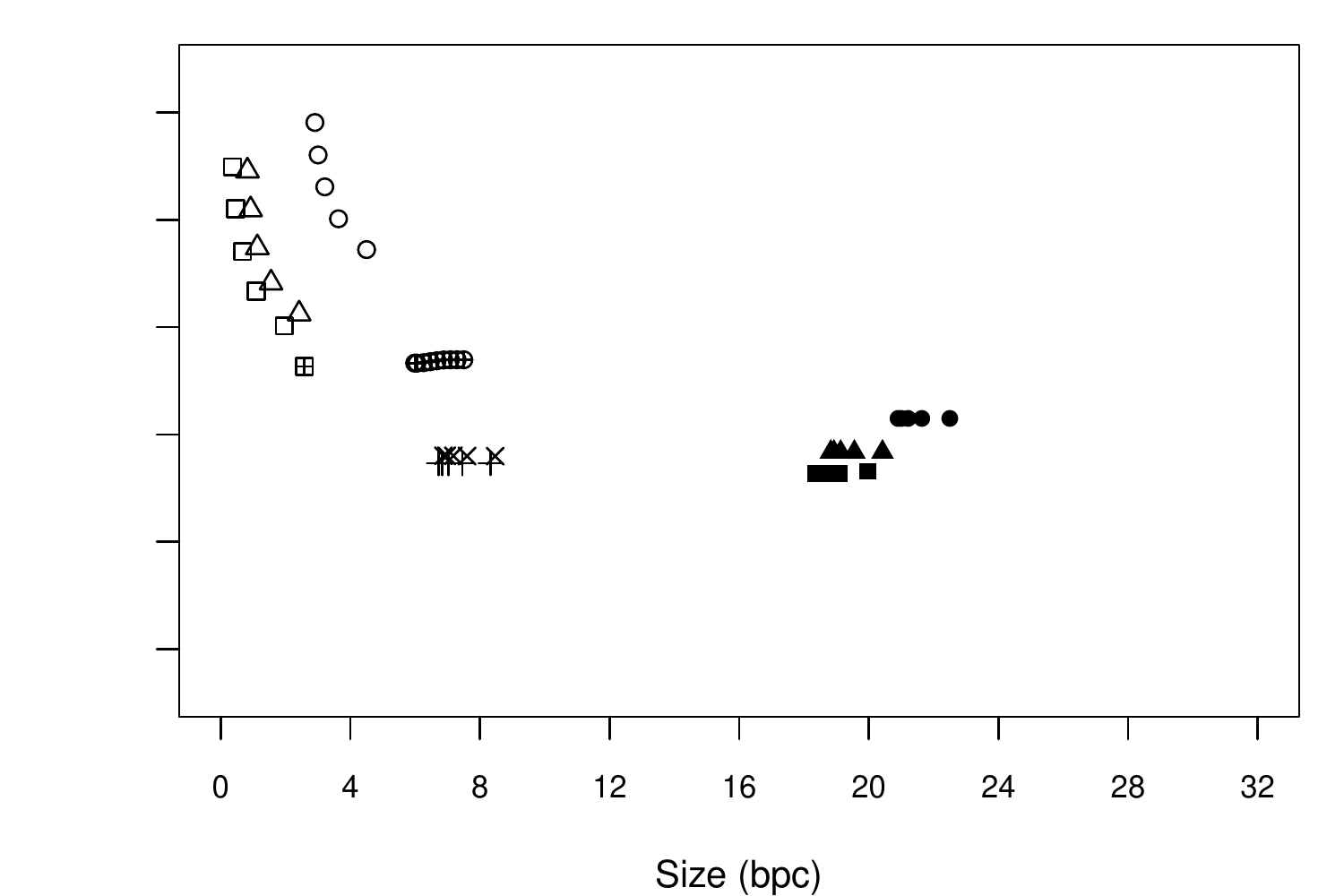}
\endminipage
\vspace{1ex}
\newline
\minipage{0.49\textwidth}
  \includegraphics[trim = 0mm 25mm 0mm 0mm, width=\linewidth]{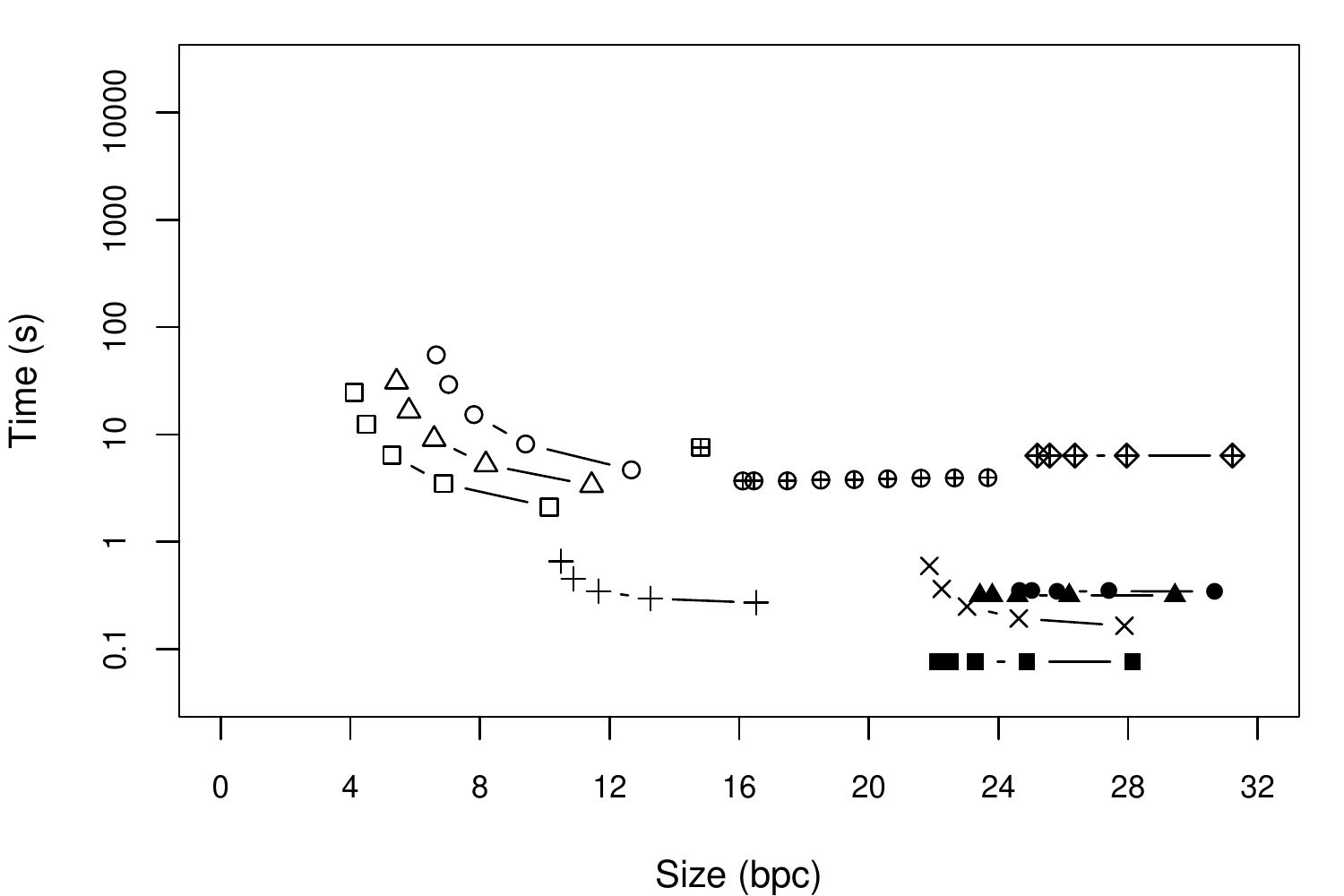}
\endminipage\hfill
\minipage{0.49\textwidth}
\endminipage
\vspace{6ex}
\caption{Document listing on small (left) and large (right) real 
collections. Total size of the index in
bits per character and time required to run the queries in seconds. From top
to bottom, \Page{}, \Revision{}, \Enwiki{}, \Influenza{}, and \Swissprot{}.}
\label{figure:doclist}
\end{figure}

\noindent {\bf Document listing with real collections.}
Figure~\ref{figure:doclist} contains the results for document listing with real collections. For most of the indexes, the time/space trade-off is based on the SA sample period. \LZ{}'s trade-off comes from a parameter specific to that structure involving RMQs (see~\cite{FN13}). \Grammar{} has no trade-off.

Of the small indexes,
\BruteL{} is usually the best choice. Thanks to the \locate() optimizations in RLCSA and the small documents, \BruteL{} beats \SadaCL{} and \SadaIL{}, which are faster in theory due to using \locate() more selectively. When more space is available, \PDLBC{} is a good choice, combining fast queries with moderate space usage. 
Of the bigger indexes, one
storing the document array explicitly is usually even faster than \PDLBC. \Grammar{} works well with \Revision{} and \Influenza{}, but becomes too large or too slow elsewhere.

\begin{figure}[p]
\minipage{0.49\textwidth}
  \includegraphics[trim = 0mm 25mm 0mm 0mm, width=\linewidth]{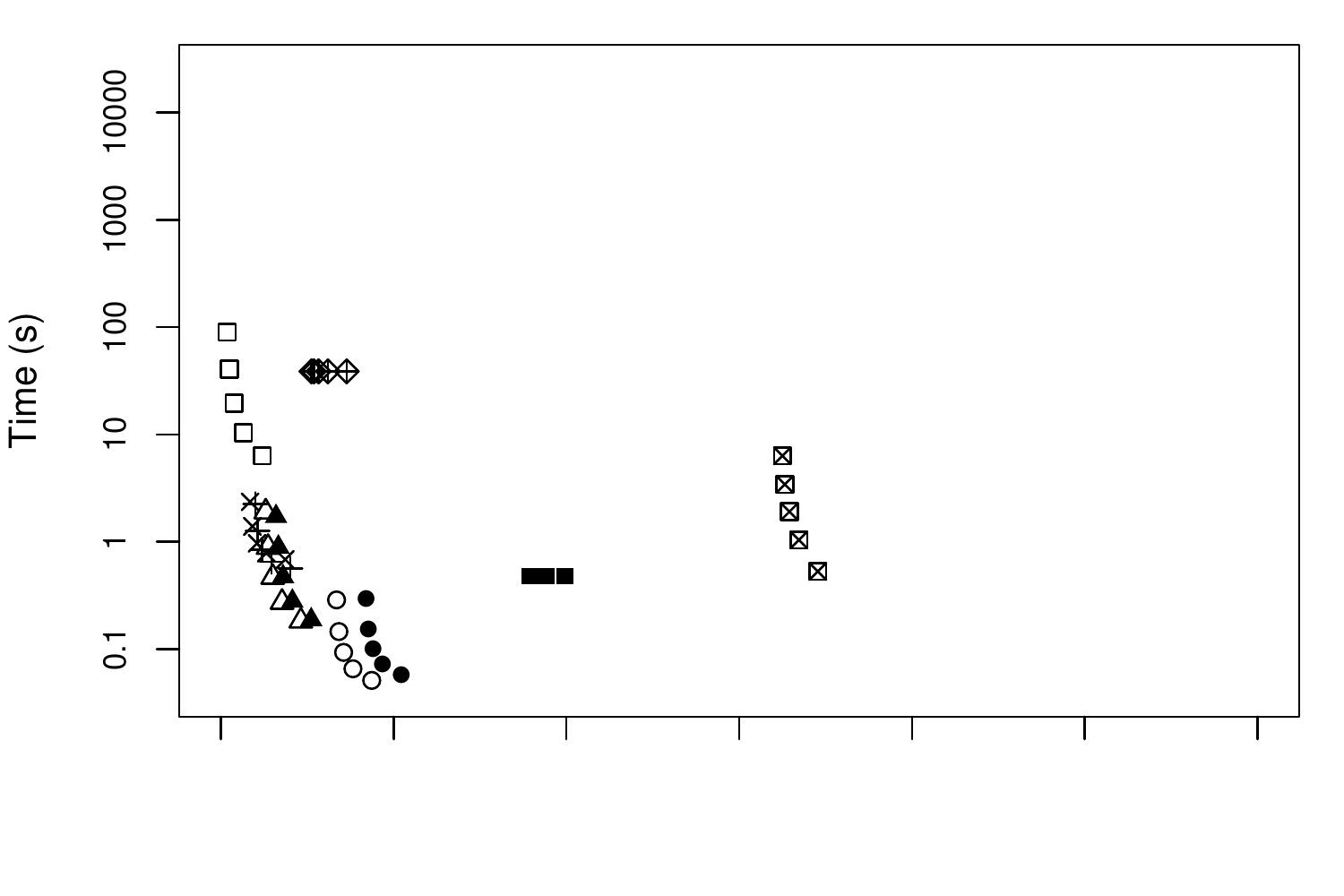}
\endminipage\hfill
\minipage{0.49\textwidth}
  \includegraphics[trim = 20mm 25mm -20mm 0mm, width=\linewidth]{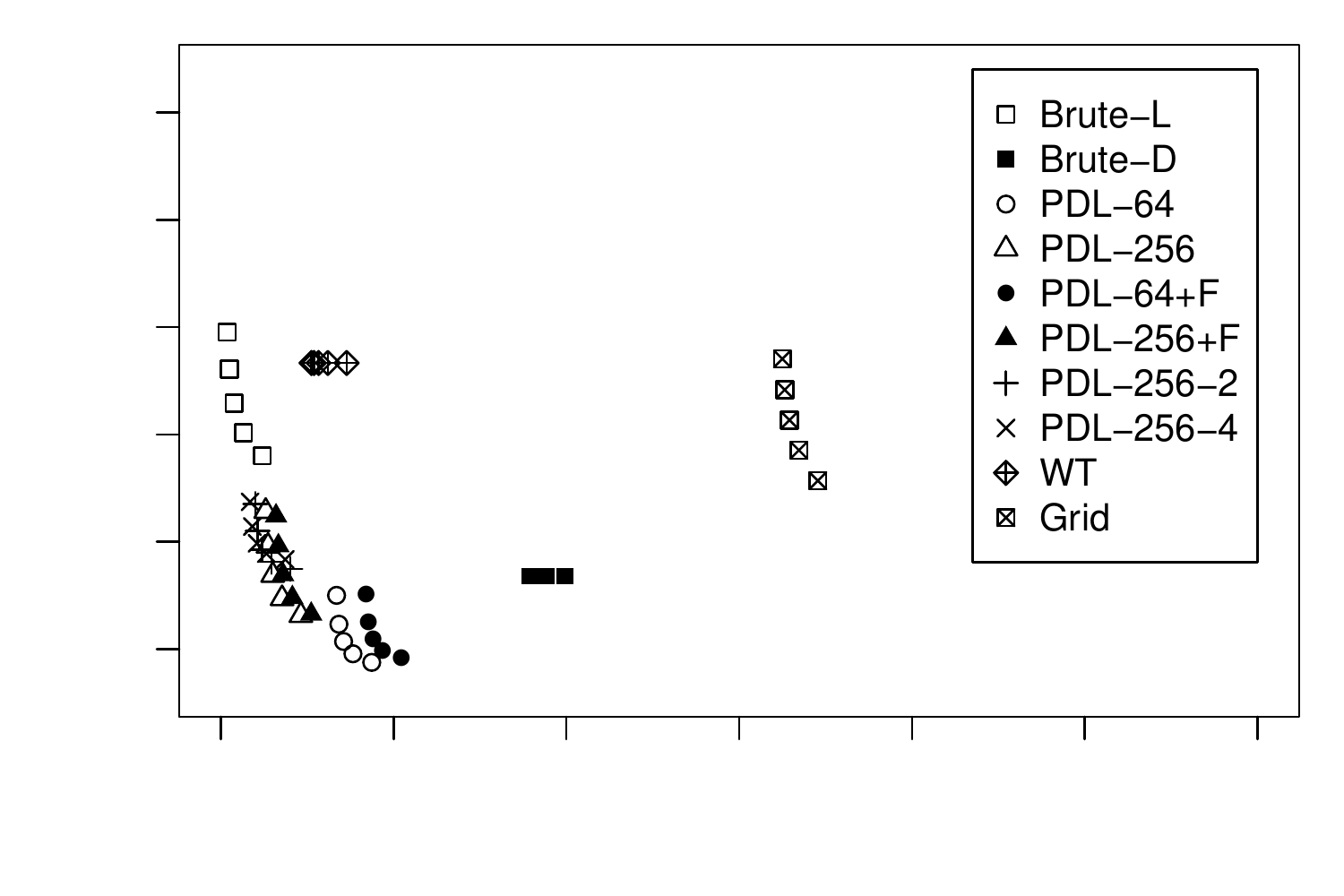}
\endminipage
\vspace{1ex}
\newline
\minipage{0.49\textwidth}
  \includegraphics[trim = 0mm 25mm 0mm 0mm, width=\linewidth]{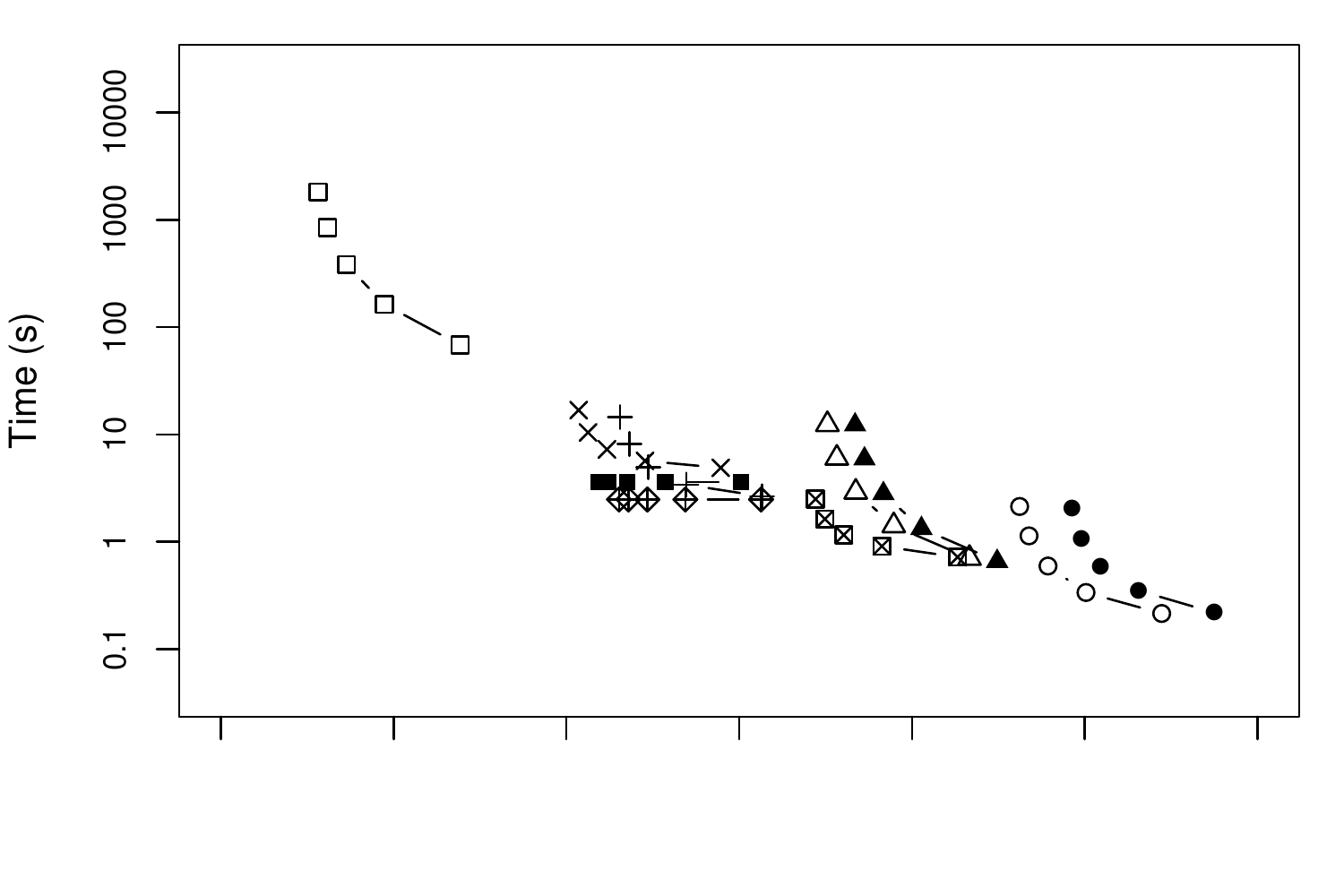}
\endminipage\hfill
\minipage{0.49\textwidth}
  \includegraphics[trim = 20mm 25mm -20mm 0mm, width=\linewidth]{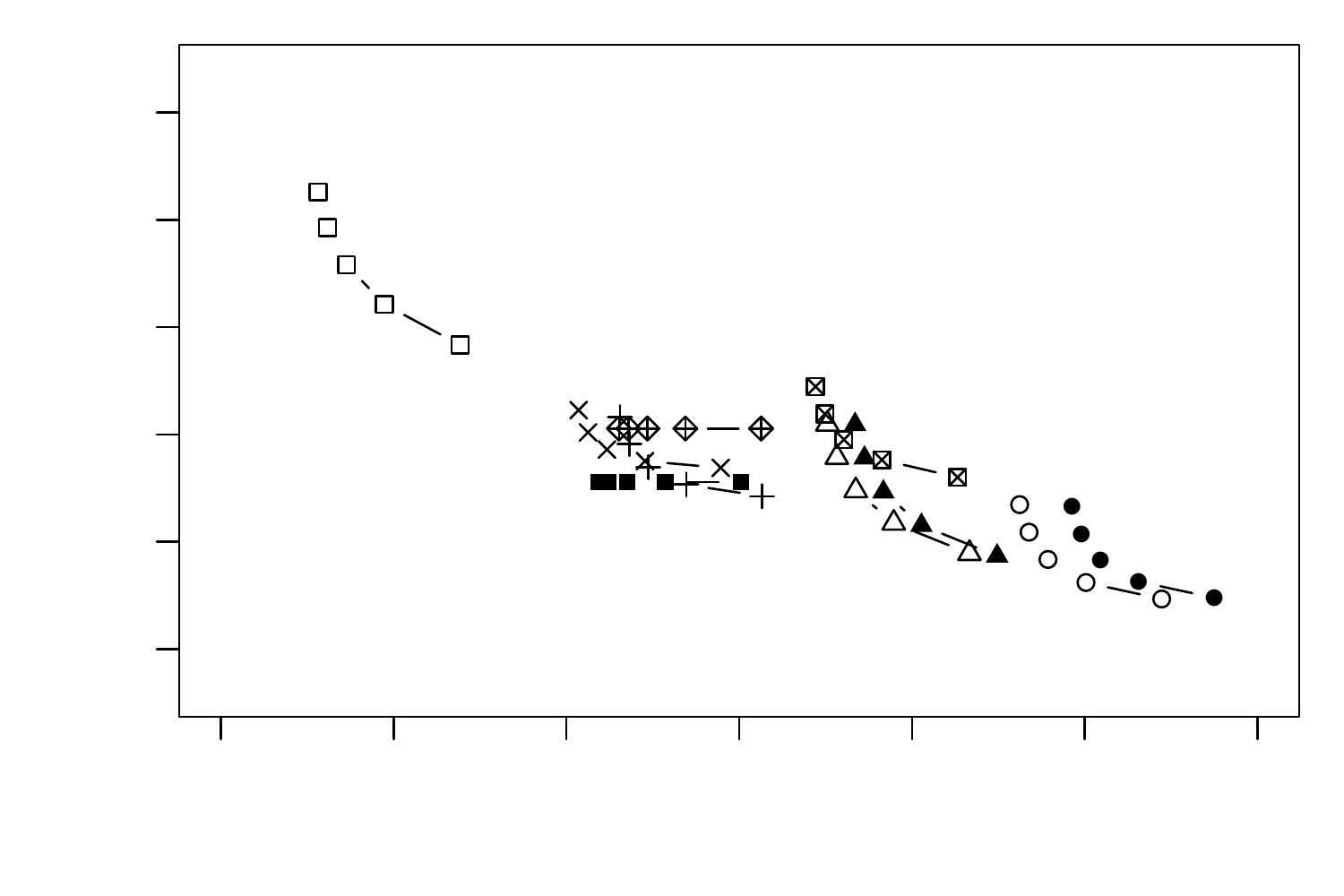}
\endminipage
\vspace{1ex}
\newline
\minipage{0.49\textwidth}
  \includegraphics[trim = 0mm 25mm 0mm 0mm, width=\linewidth]{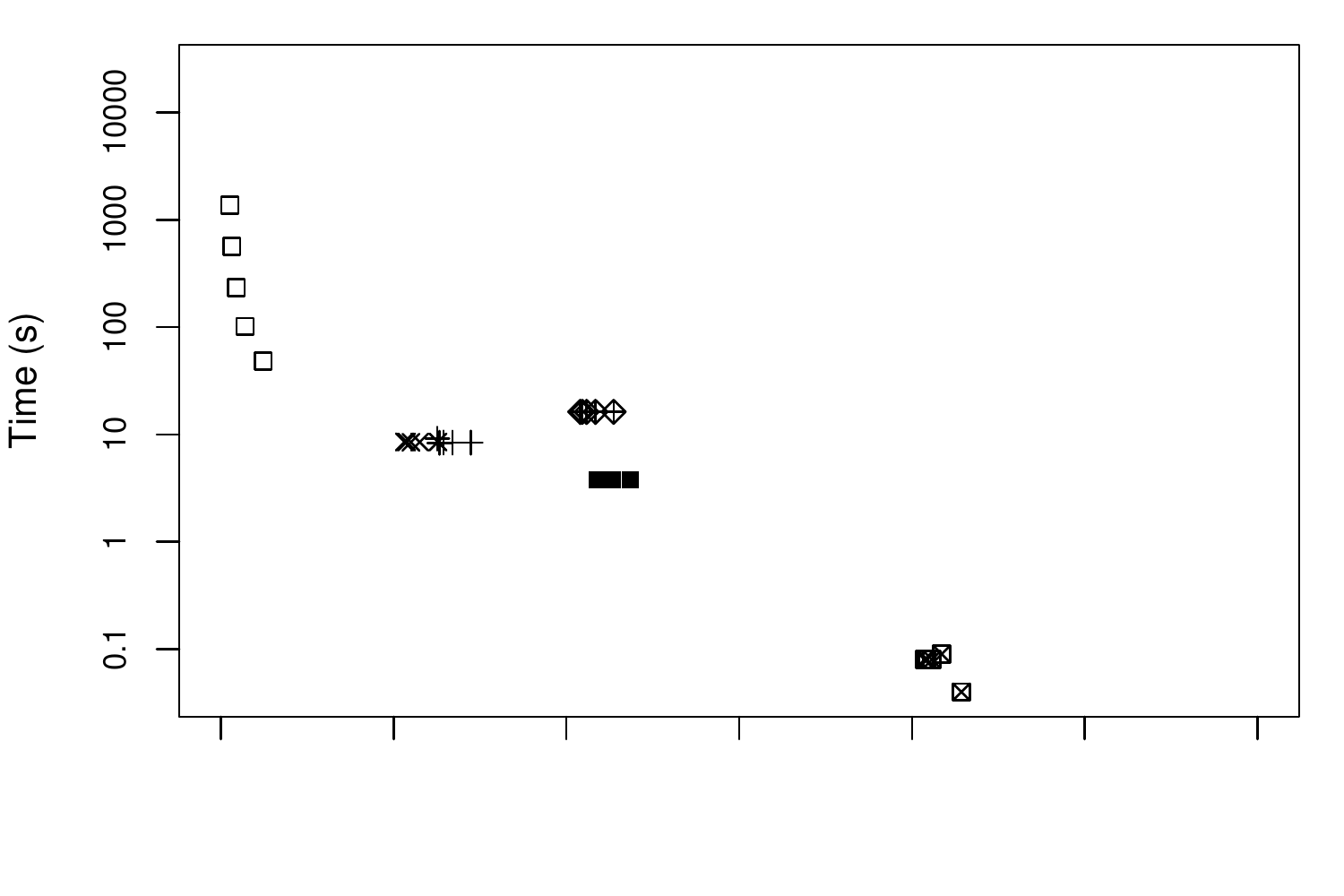}
\endminipage\hfill
\minipage{0.49\textwidth}
  \includegraphics[trim = 20mm 25mm -20mm 0mm, width=\linewidth]{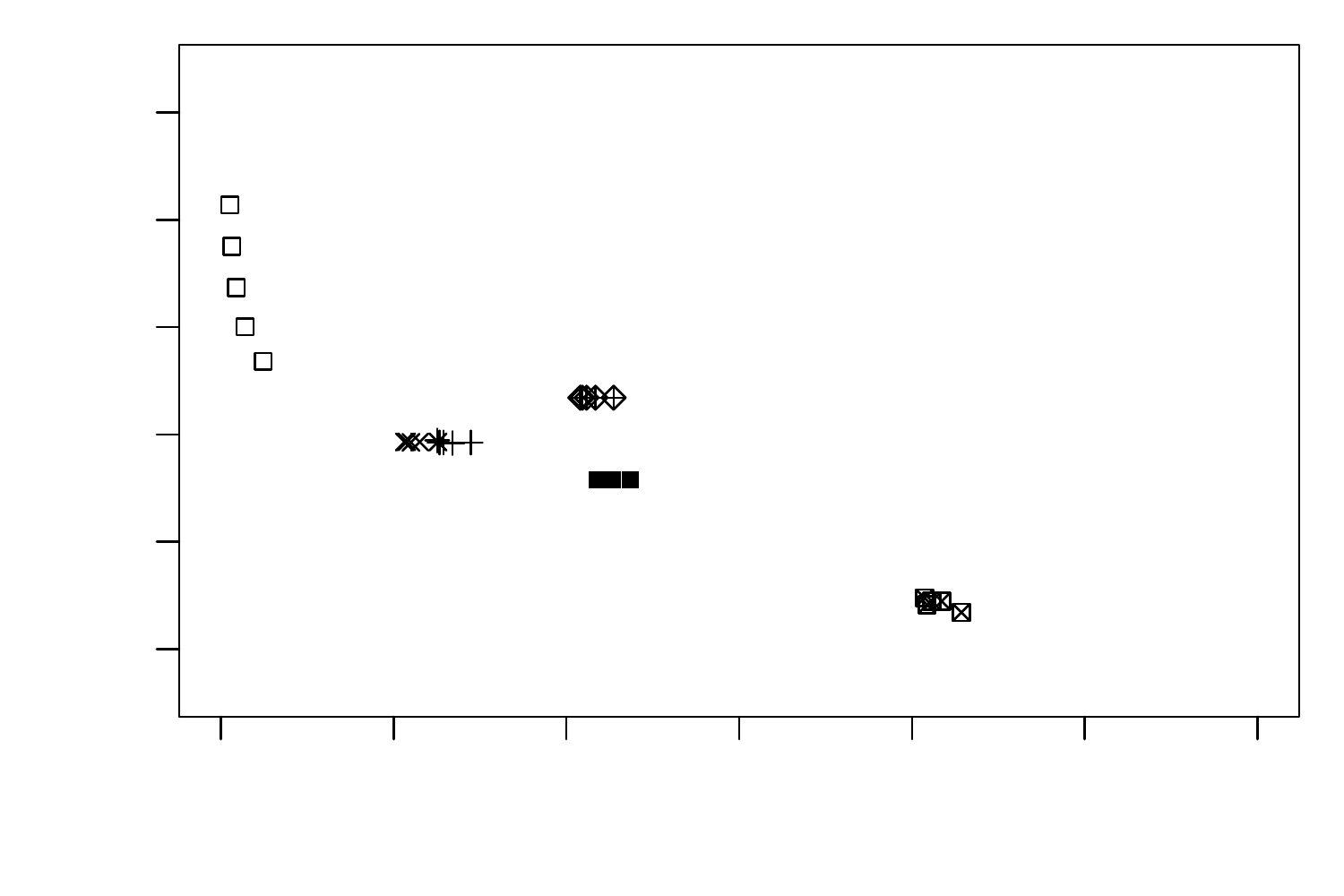}
\endminipage
\vspace{1ex}
\newline
\minipage{0.49\textwidth}
  \includegraphics[trim = 0mm 25mm 0mm 0mm, width=\linewidth]{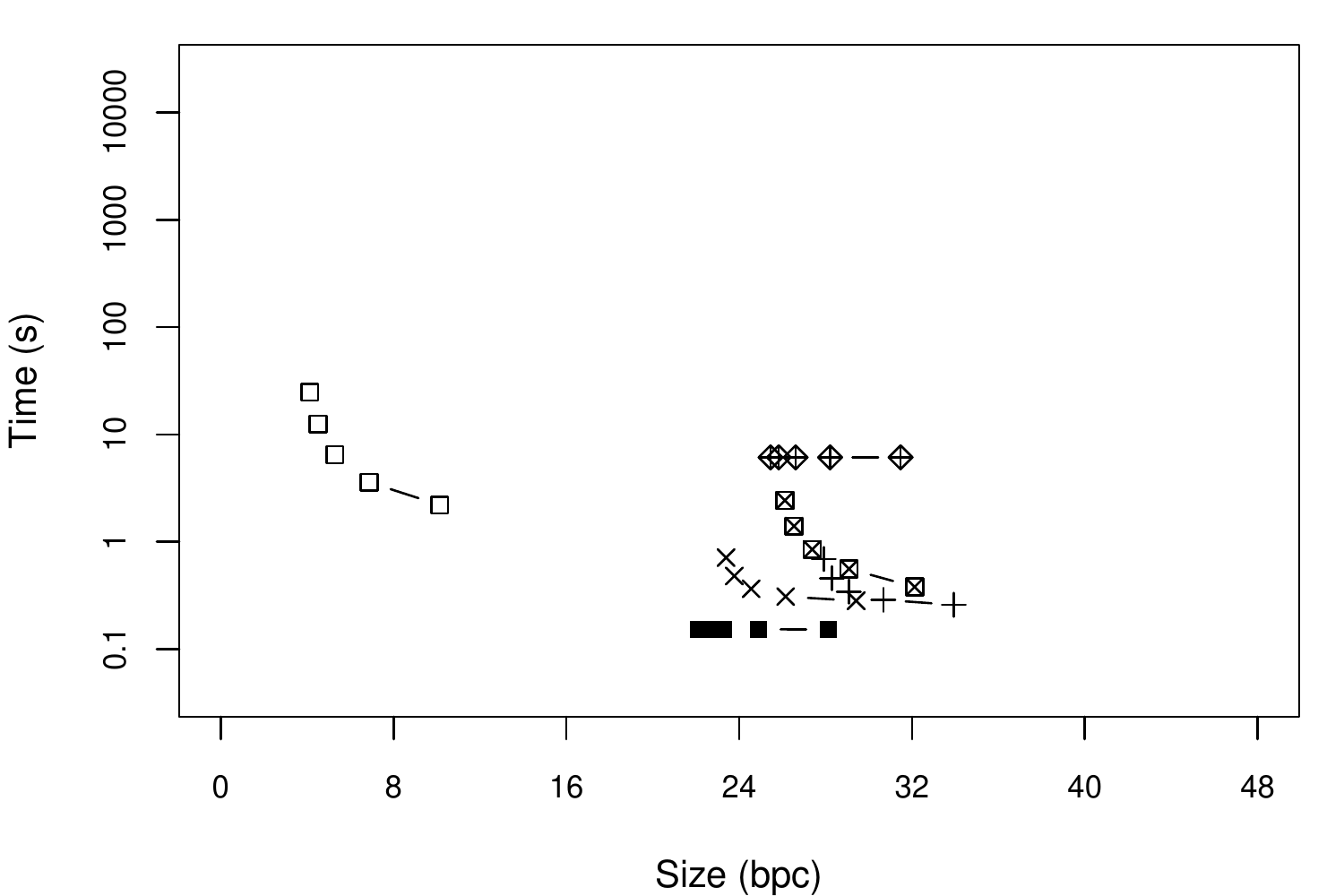}
\endminipage\hfill
\minipage{0.49\textwidth}
  \includegraphics[trim = 20mm 25mm -20mm 0mm, width=\linewidth]{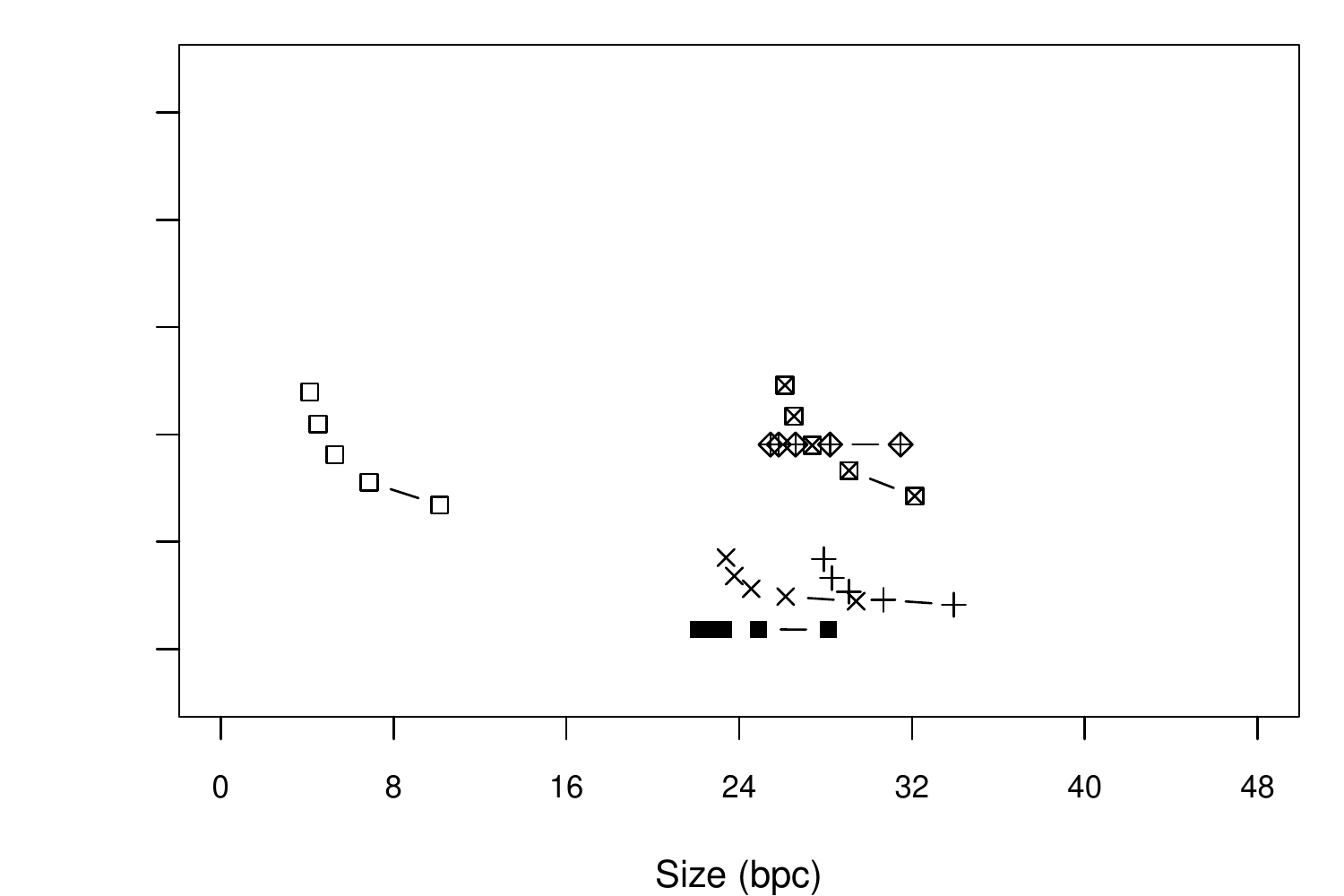}
\endminipage
\vspace{6ex}
\caption{Top-$k$ document retrieval with $k=10$ (left) and $k=100$ (right) on small real collections. Total size of the index in
bits per character and time required to run the queries in seconds. From top
to bottom, \Revision{}, \Enwiki{}, \Influenza{}, and \Swissprot{}.
\Page{} is left out due to the low number of documents in 
that collection.}
\label{figure:topk-small}
\end{figure}

\begin{figure}[t]
\minipage{0.49\textwidth}
  \includegraphics[trim = 0mm 25mm 0mm 0mm, width=\linewidth]{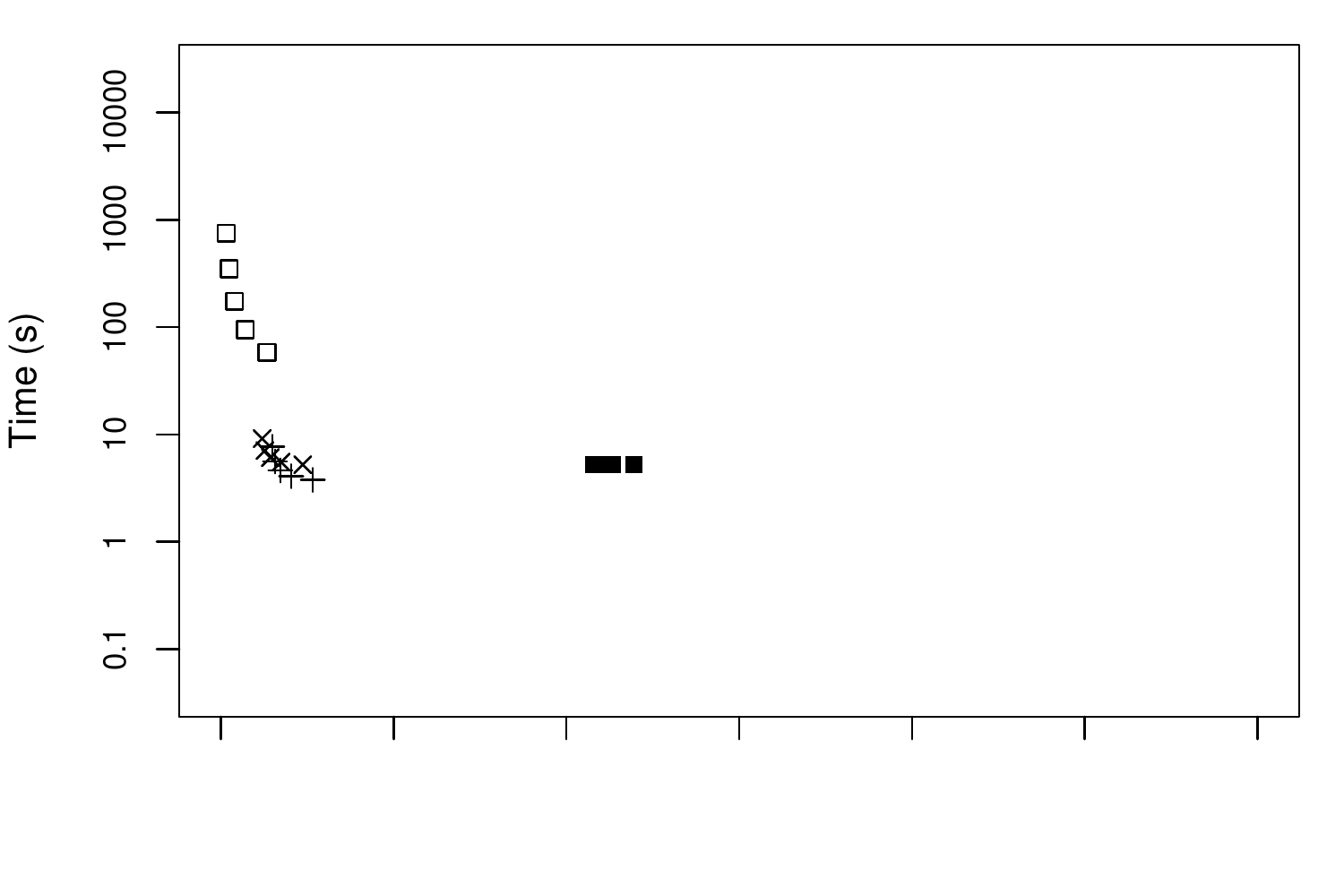}
\endminipage\hfill
\minipage{0.49\textwidth}
  \includegraphics[trim = 20mm 25mm -20mm 0mm, width=\linewidth]{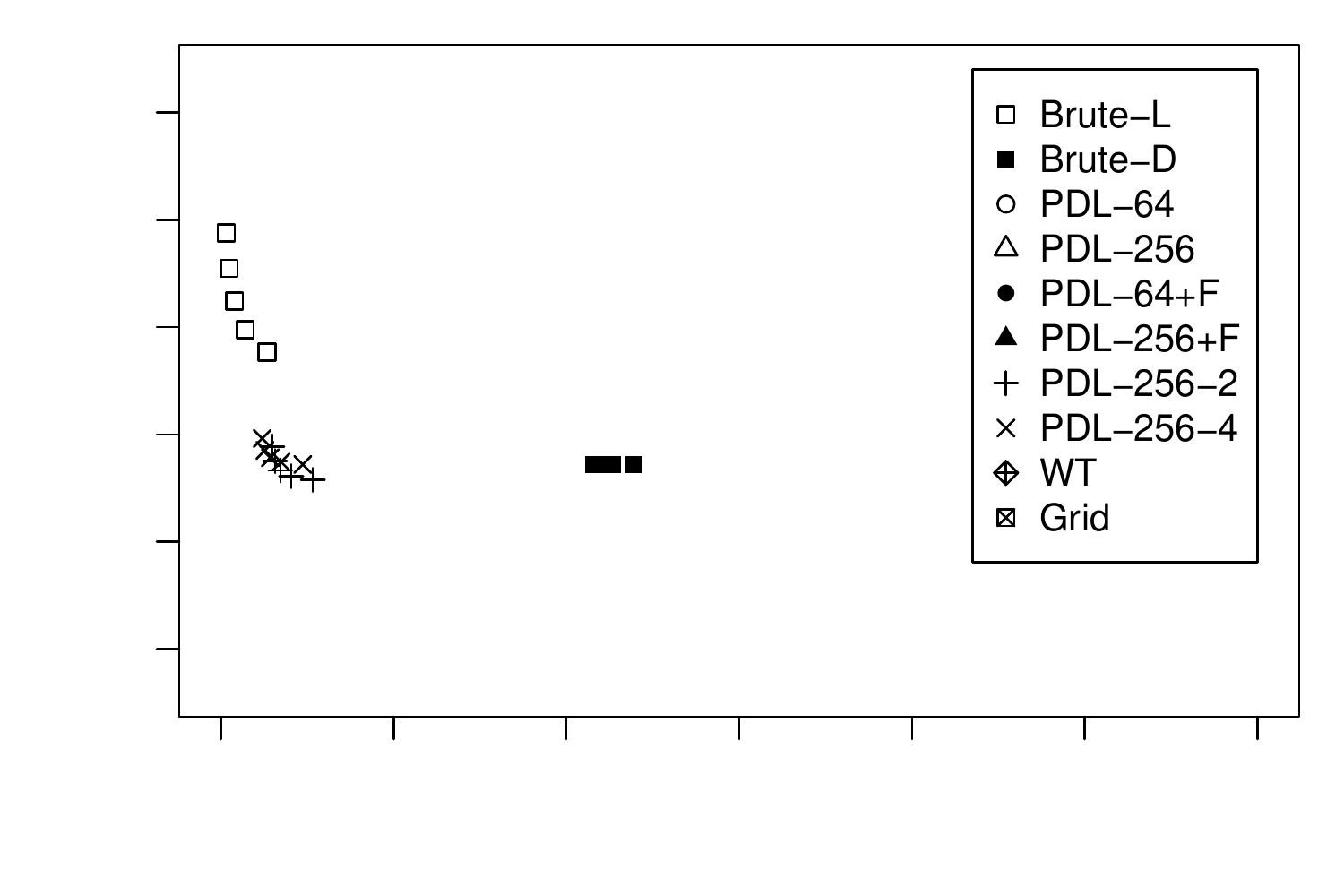}
\endminipage
\vspace{1ex}
\newline
\minipage{0.49\textwidth}
  \includegraphics[trim = 0mm 25mm 0mm 0mm, width=\linewidth]{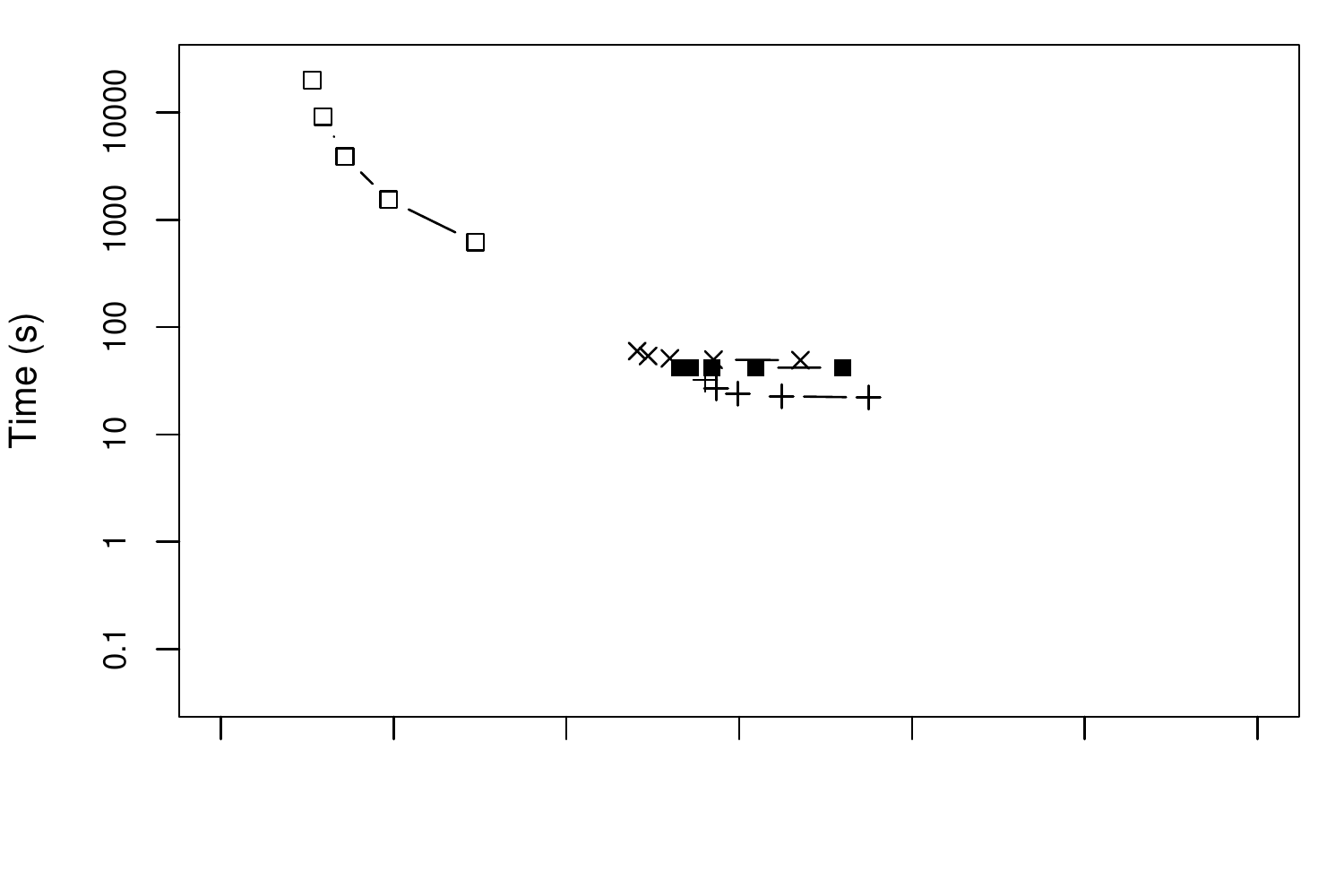}
\endminipage\hfill
\minipage{0.49\textwidth}
  \includegraphics[trim = 20mm 25mm -20mm 0mm, width=\linewidth]{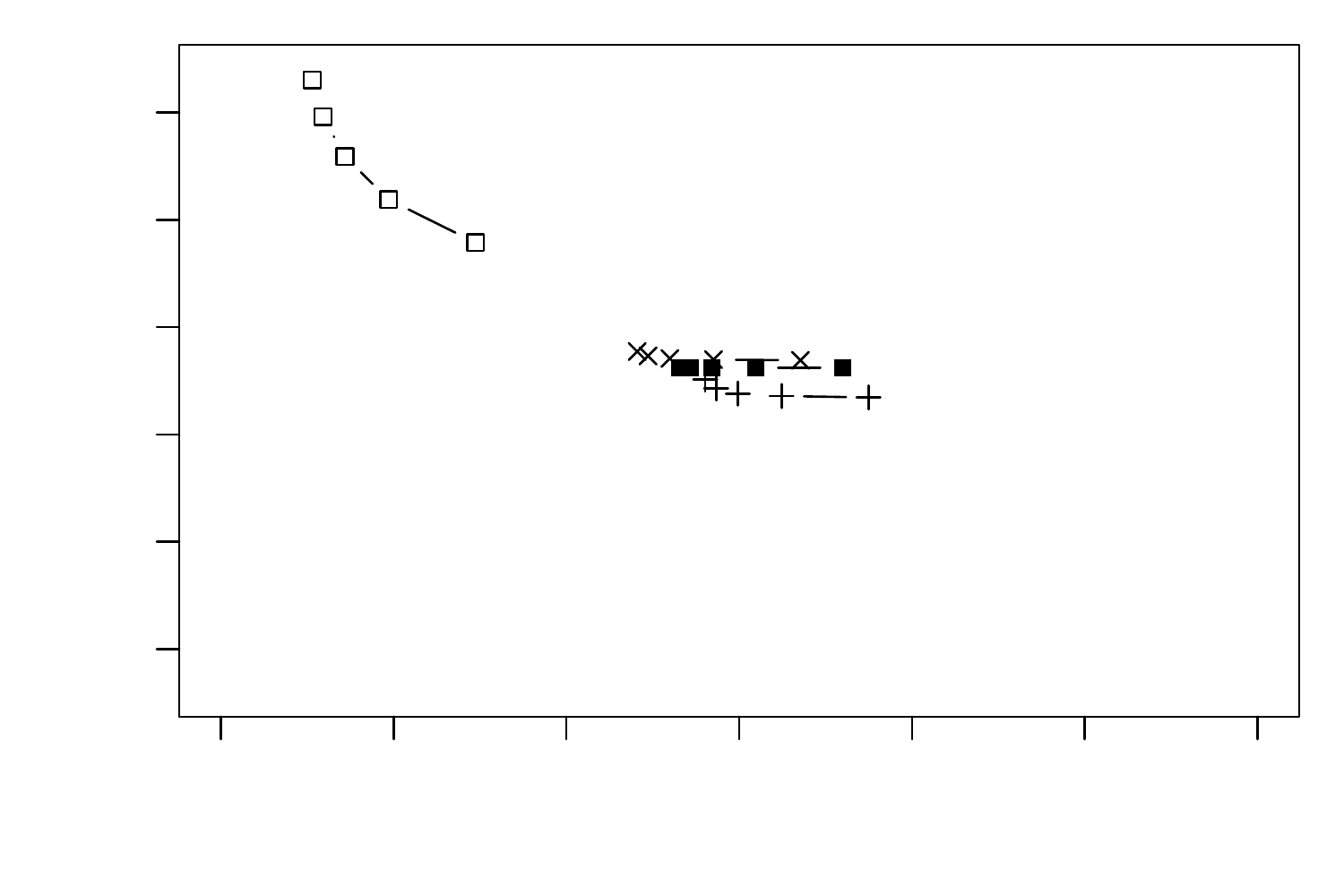}
\endminipage
\vspace{1ex}
\newline
\minipage{0.49\textwidth}
  \includegraphics[trim = 0mm 25mm 0mm 0mm, width=\linewidth]{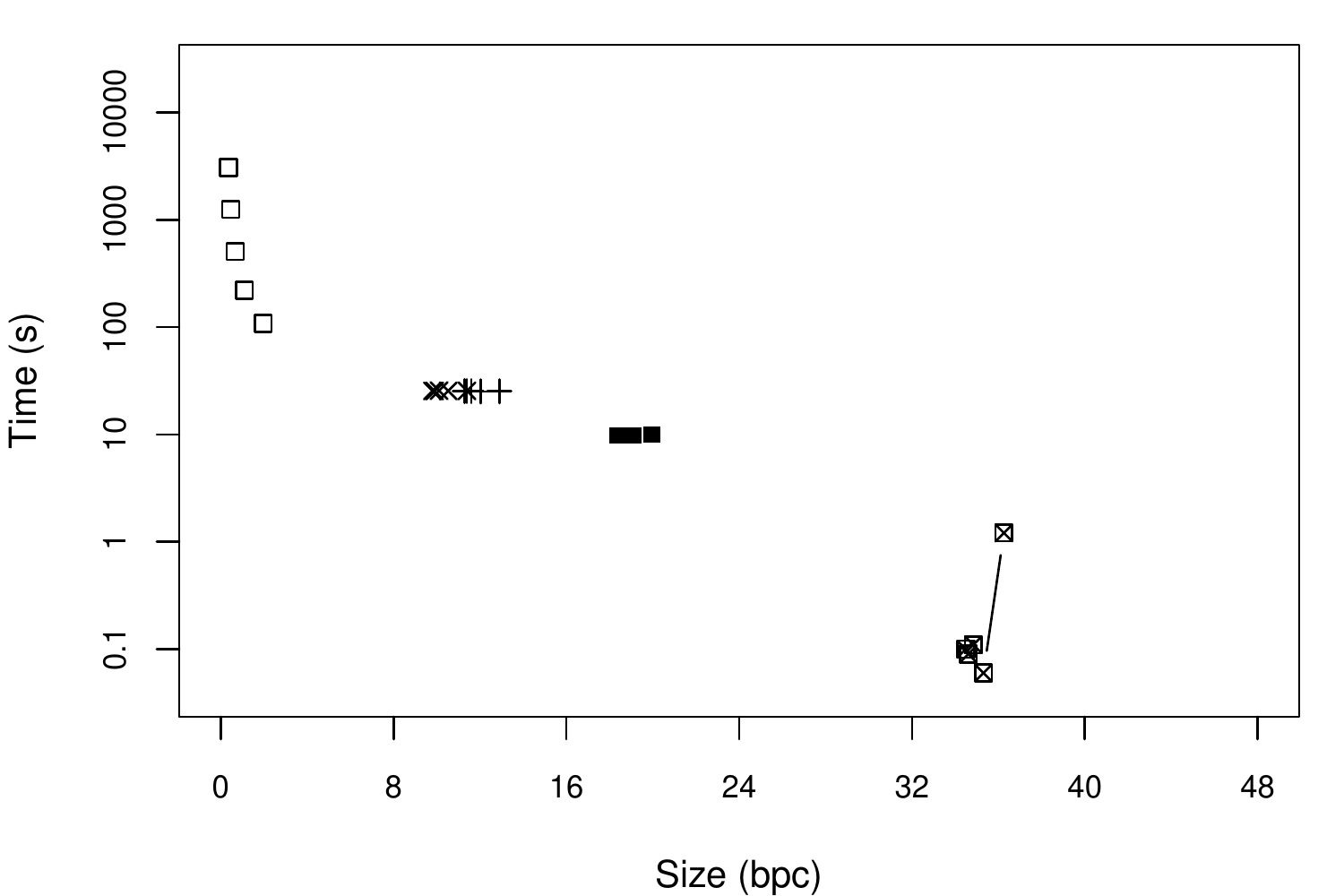}
\endminipage\hfill
\minipage{0.49\textwidth}
  \includegraphics[trim = 20mm 25mm -20mm 0mm, width=\linewidth]{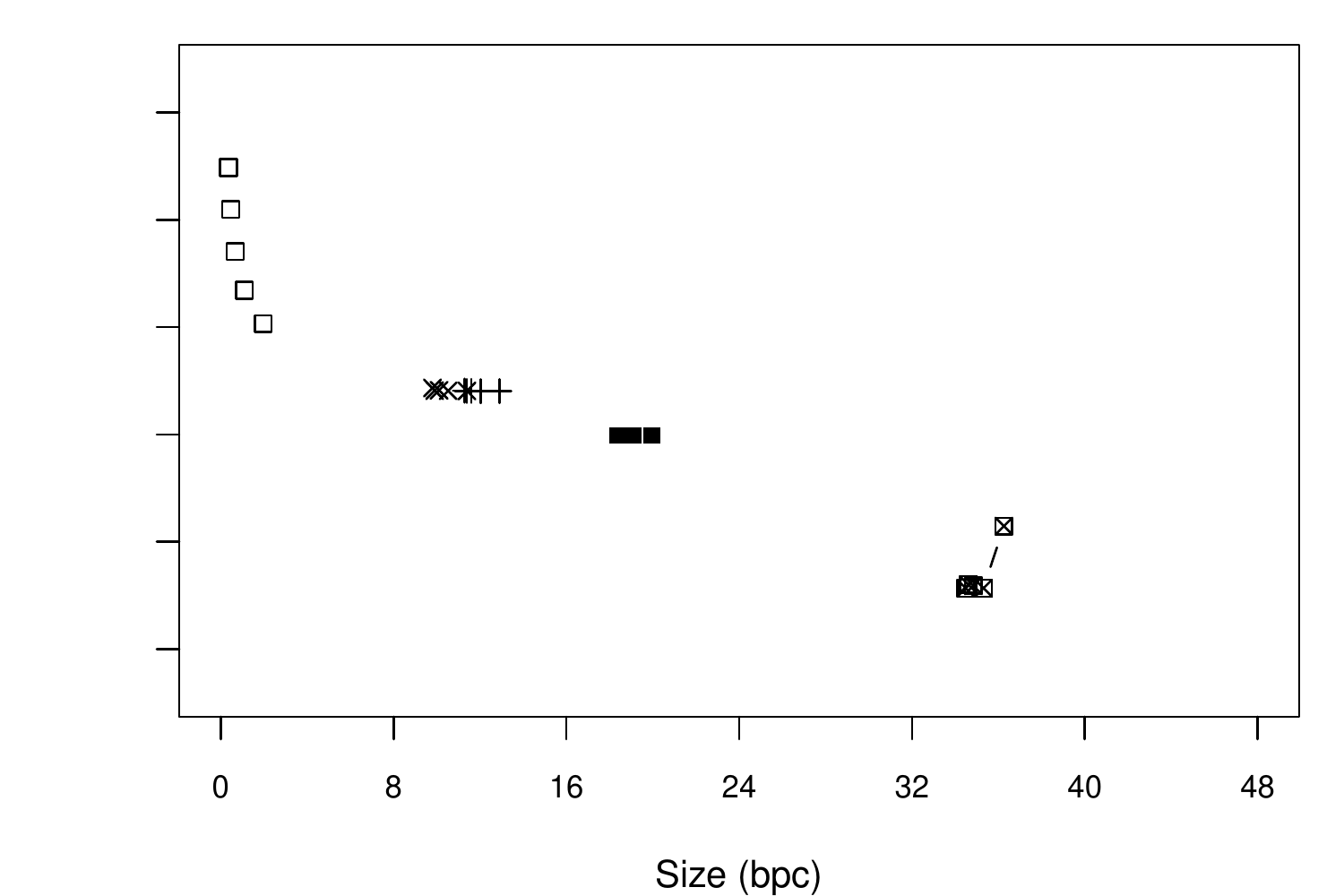}
\endminipage
\vspace{6ex}
\caption{Top-$k$ document retrieval with $k=10$ (left) and $k=100$ (right) on large real collections. Total size of the index in
bits per character and time required to run the queries in seconds. From top
to bottom, \Revision{}, \Enwiki{}, and \Influenza{}.
\Page{} is left out due to the low number of documents in 
that collection.}
\label{figure:topk-large}
\end{figure}

\noindent {\bf Top-$k$ document retrieval.}
Results for top-$k$ document retrieval on real collections are shown in Figures~\ref{figure:topk-small} and \ref{figure:topk-large}.
Time/space trade-offs are again based on the suffix array sample period, while \PDL{} also uses other parameters (see Section~\ref{section:pdl}). We could not build \PDL{} with $\beta = 0$ for \Influenza{} or the large collections, as the total size of the stored sets was more than $2^{32}$, which was too much for our \RePair{} compressor. \WT{} was only built for the small collections, while \Grid{} construction used too much memory on the larger Wikipedia collections.

On \Revision, \PDL{} dominates the other solutions.
On \Enwiki, both \WT{} and \Grid{} have good trade-offs with $k=10$, while \BruteD{} and \PDL{} beat them with $k=100$. On \Influenza{}, some \PDL{} variants, \BruteD{}, and \Grid{} all offer good trade-offs. On \Swissprot{}, the brute-force algorithms win clearly. \PDL{} with $\beta=0$ is faster, but requires far too much space ($60$-$70$ bpc --- off the chart).

\begin{figure}
\minipage{0.49\textwidth}
  \includegraphics[trim = 0mm 25mm 0mm 0mm, width=\linewidth]{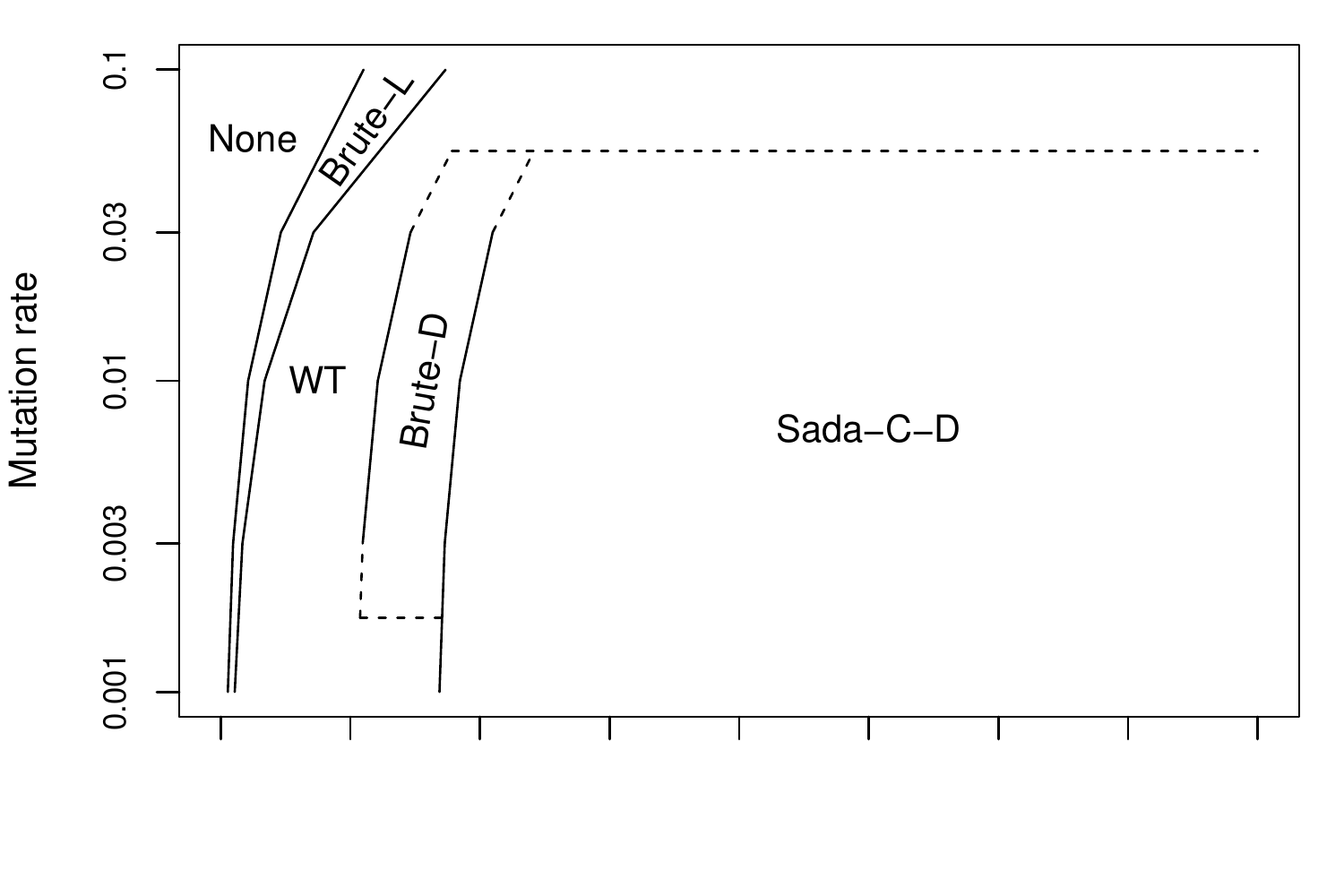}
\endminipage\hfill
\minipage{0.49\textwidth}
  \includegraphics[trim = 20mm 25mm -20mm 0mm, width=\linewidth]{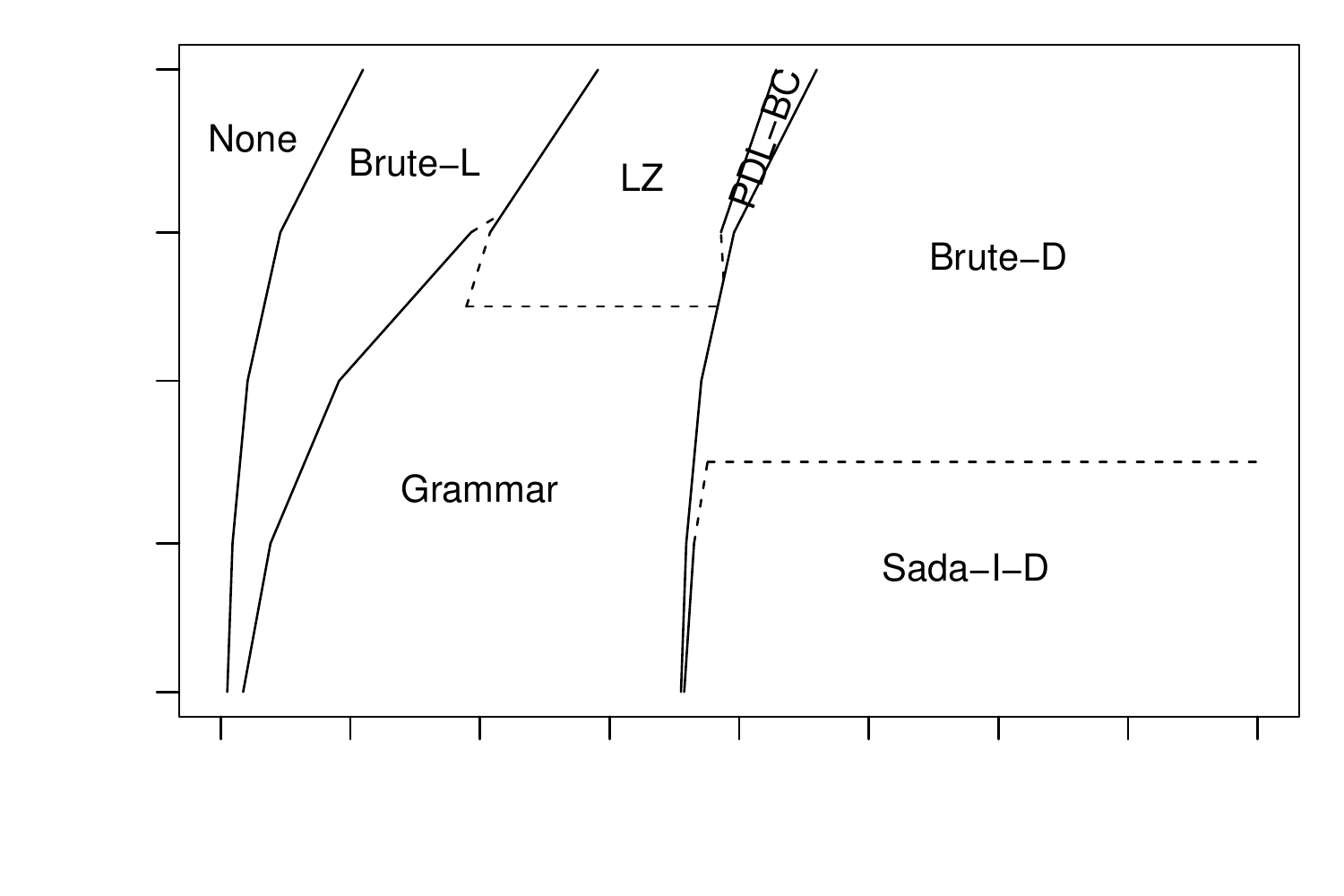}
\endminipage
\vspace{1ex}
\newline
\minipage{0.49\textwidth}
  \includegraphics[trim = 0mm 25mm 0mm 0mm, width=\linewidth]{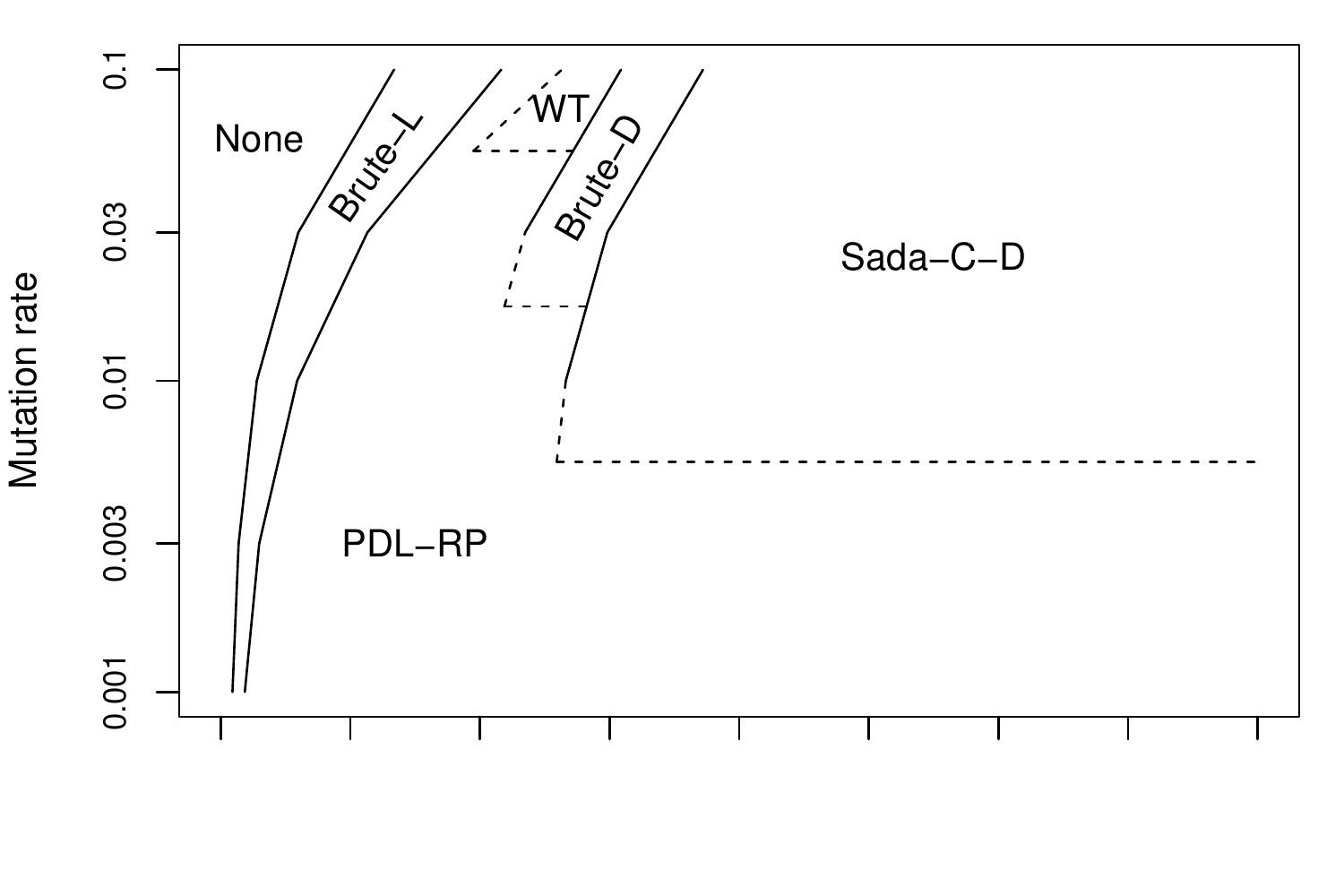}
\endminipage\hfill
\minipage{0.49\textwidth}
  \includegraphics[trim = 20mm 25mm -20mm 0mm, width=\linewidth]{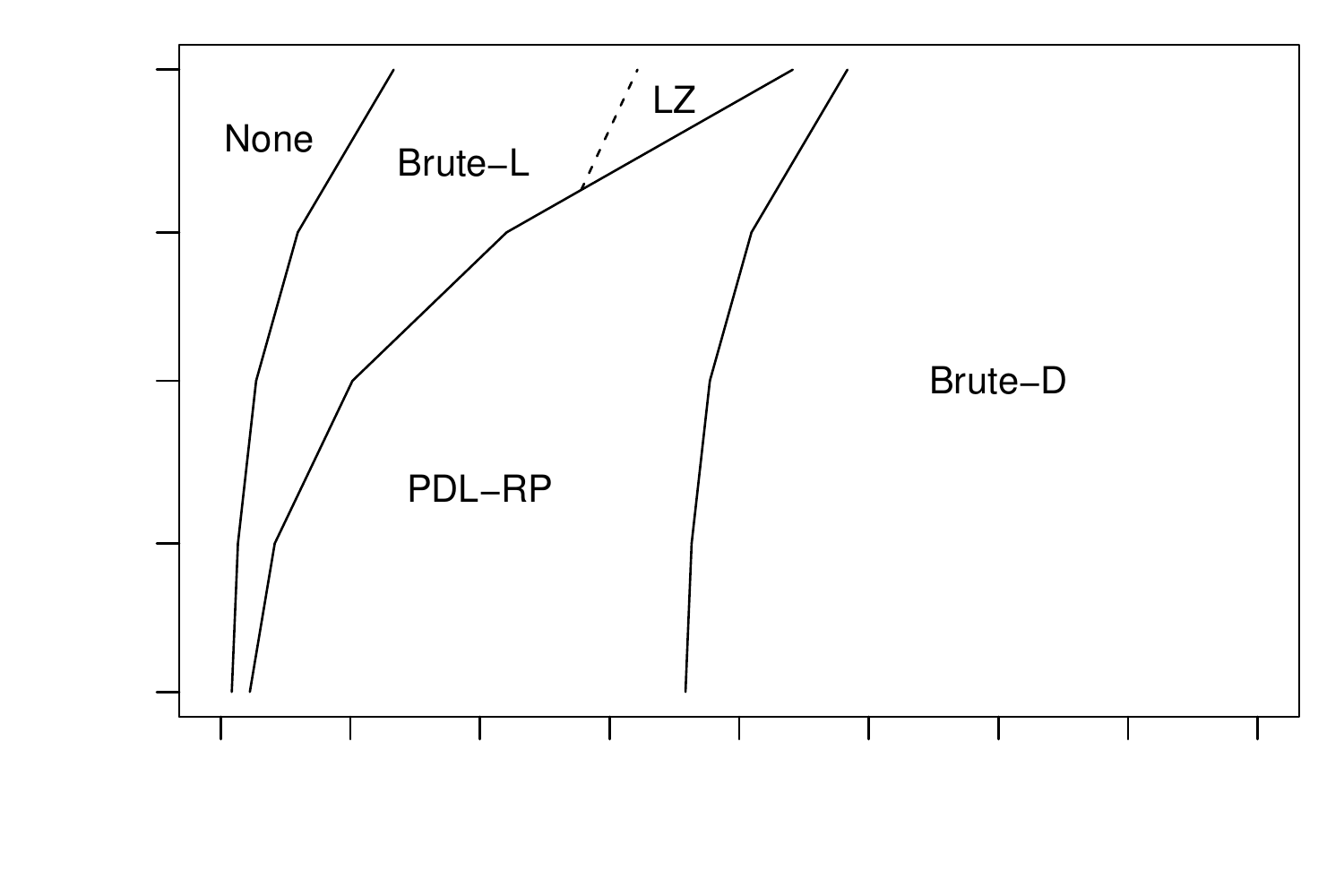}
\endminipage
\vspace{1ex}
\newline
\minipage{0.49\textwidth}
  \includegraphics[trim = 0mm 25mm 0mm 0mm, width=\linewidth]{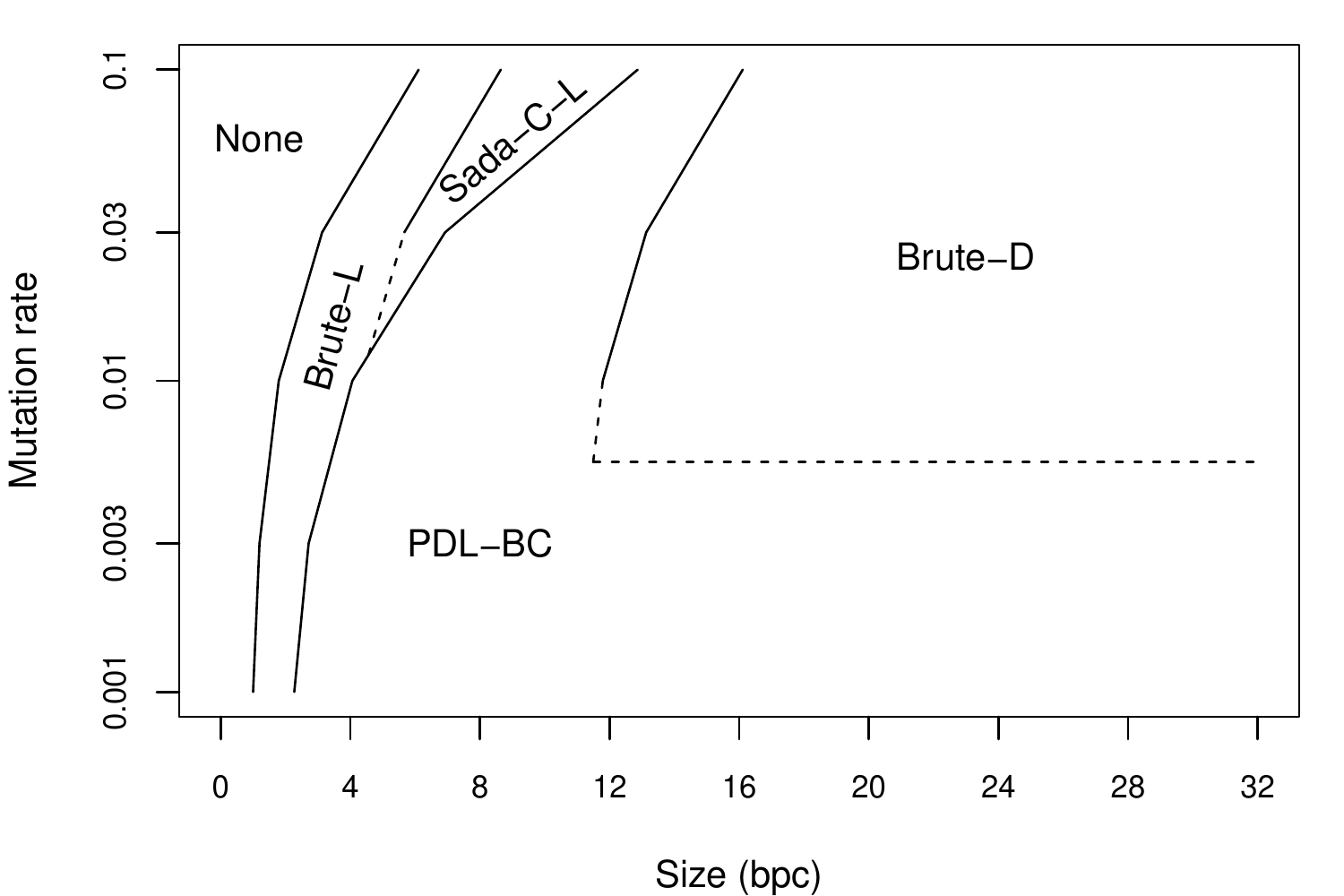}
\endminipage\hfill
\minipage{0.49\textwidth}
  \includegraphics[trim = 20mm 25mm -20mm 0mm, width=\linewidth]{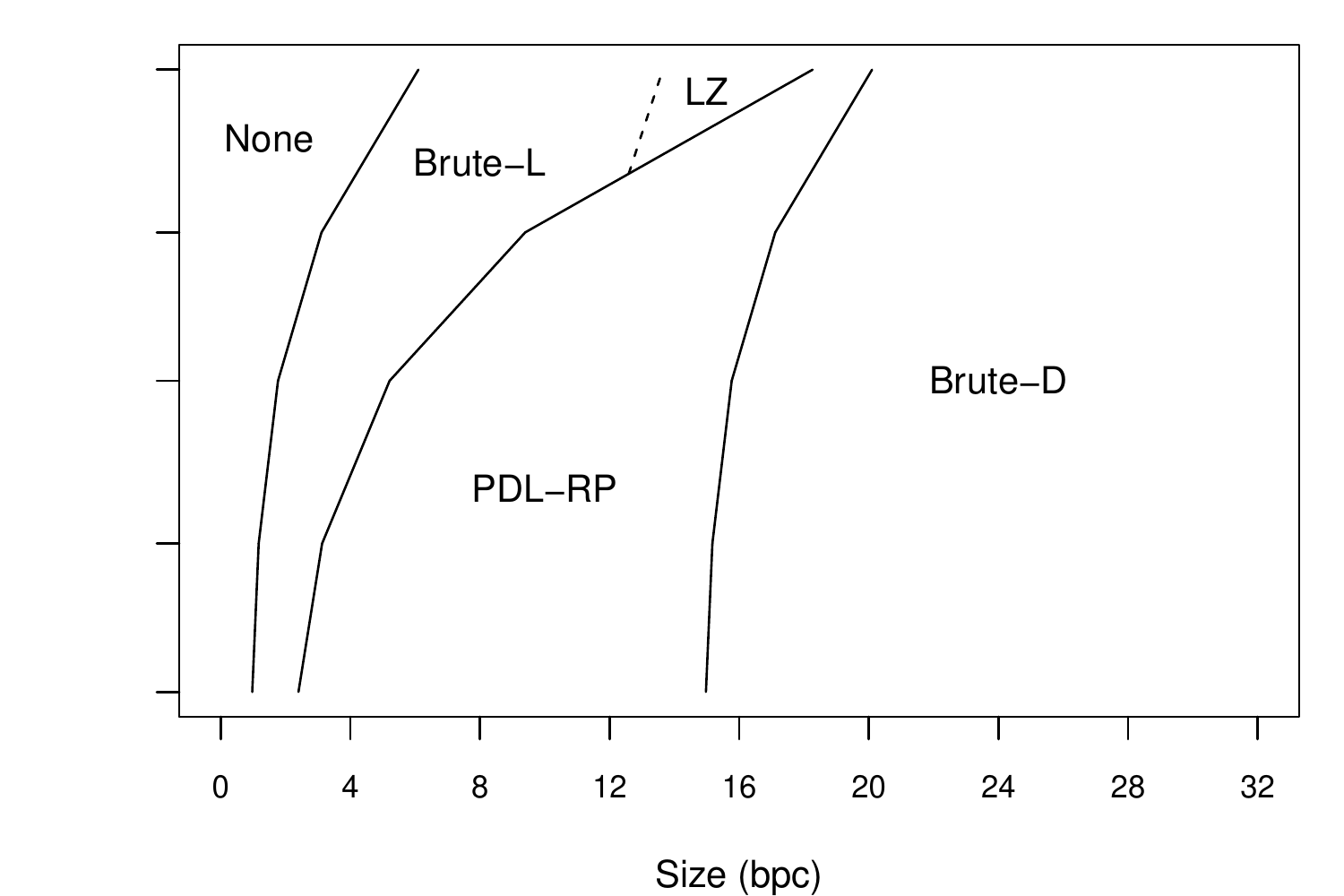}
\endminipage
\vspace{7ex}
\minipage{0.49\textwidth}
  \includegraphics[trim = 0mm 25mm 0mm 0mm, width=\linewidth]{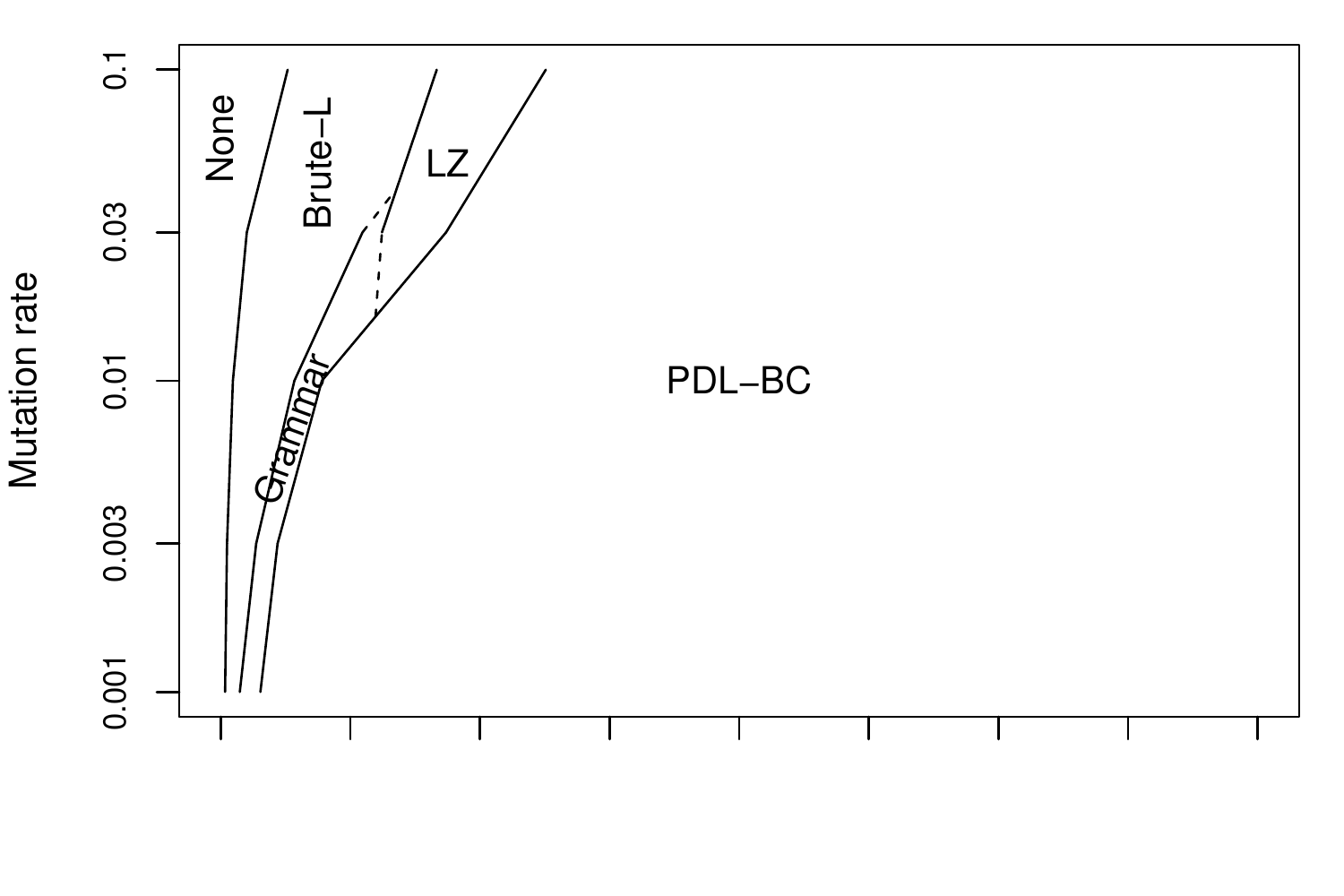}
\endminipage\hfill
\minipage{0.49\textwidth}
  \includegraphics[trim = 20mm 25mm -20mm 0mm, width=\linewidth]{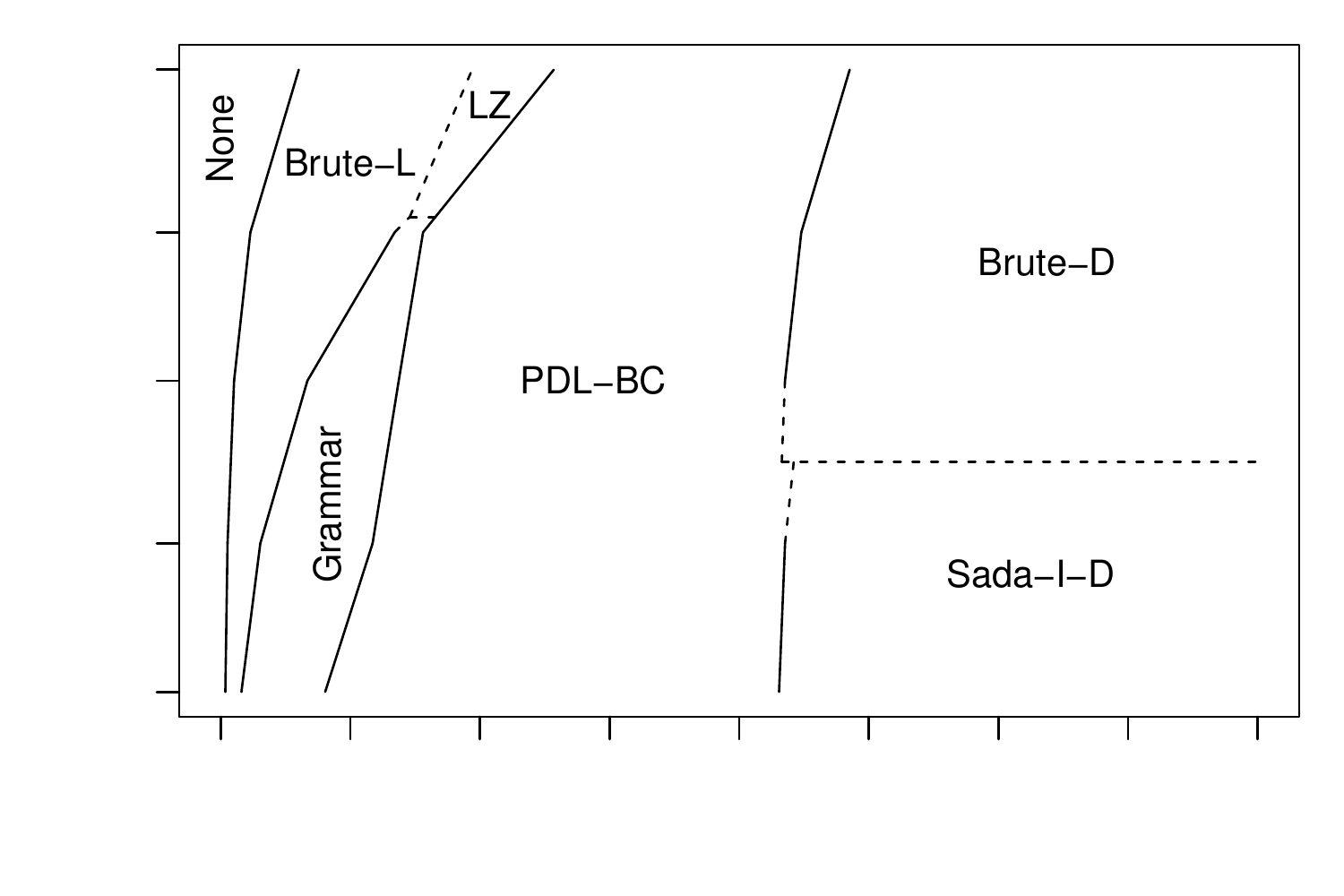}
\endminipage
\vspace{1ex}
\newline
\minipage{0.49\textwidth}
  \includegraphics[trim = 0mm 25mm 0mm 0mm, width=\linewidth]{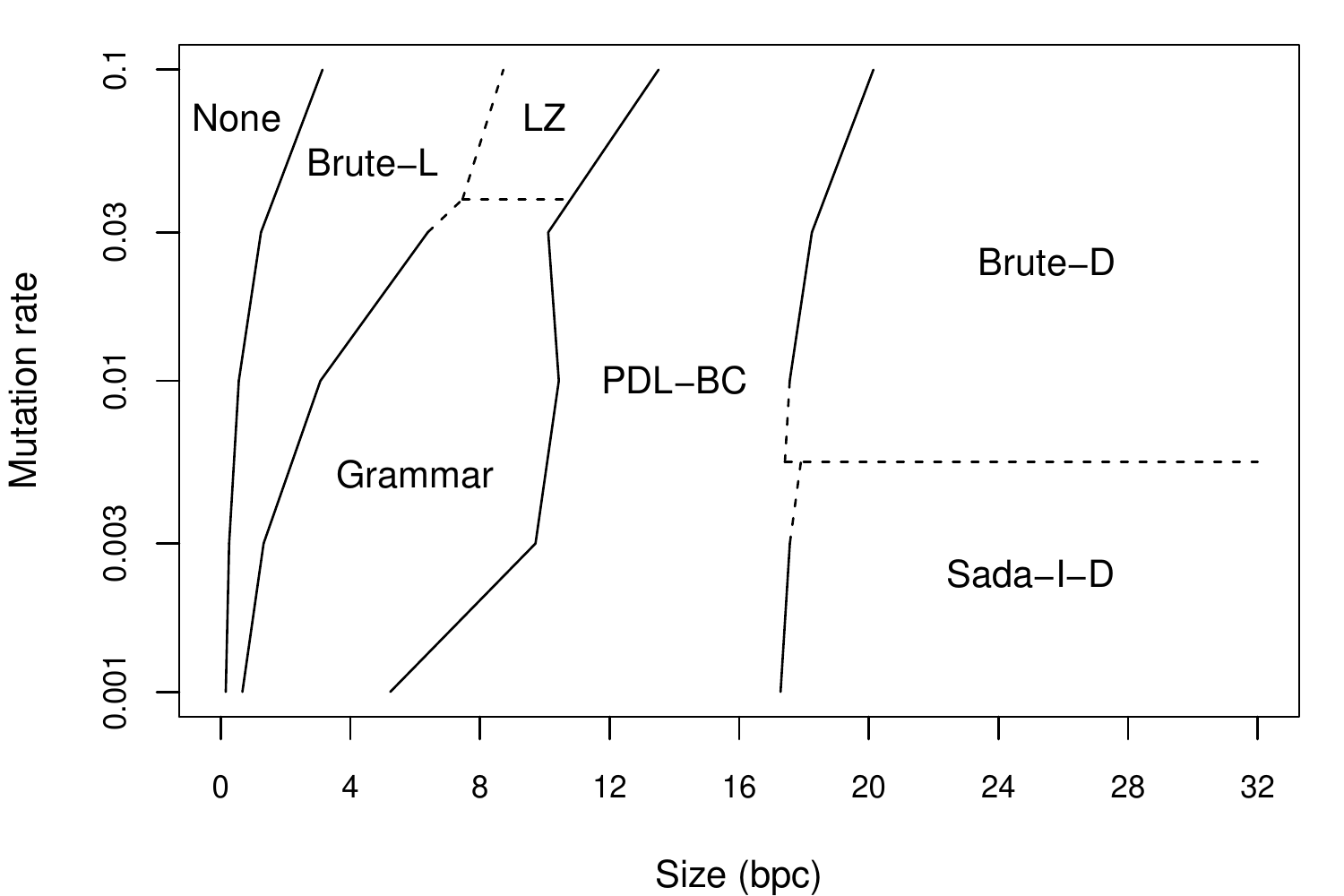}
\endminipage\hfill
\minipage{0.49\textwidth}
  \includegraphics[trim = 20mm 25mm -20mm 0mm, width=\linewidth]{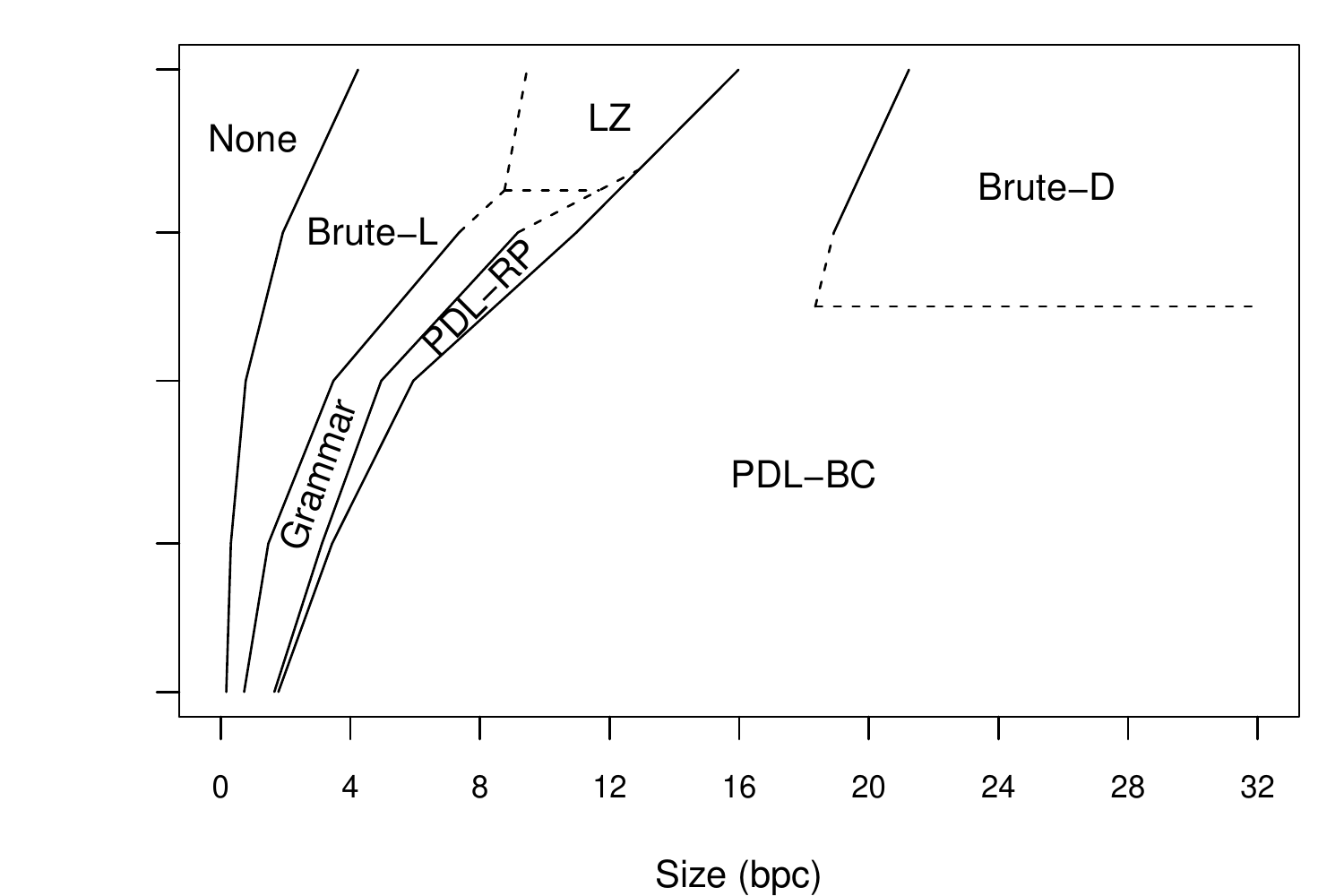}
\endminipage
\vspace{6ex}
\caption{Document listing with synthetic collections. The fastest solution for a given size in bits per character and a mutation rate. Top group: from top to bottom $10$, $100$, and $1000$ base documents with \Concat{} (left) and \Version{} (right). Bottom group: \DNA{} with $1$ (top left), $10$ (top right), $100$ (bottom left), and $1000$ (bottom right) base documents. \textsf{None} denotes that no solution can achieve that size.}\label{figure:synthetic}
\end{figure}

\noindent{\bf Document listing with synthetic collections.}
Figure~\ref{figure:synthetic} shows our document listing results with synthetic collections. Due to the large number of collections, the results for a given collection type and number of base documents are combined in a single plot, showing the fastest algorithm for a given amount of space and a mutation rate. Solid lines connect measurements that are the fastest for their size, while dashed lines are rough interpolations.

The plots were simplified in two ways. Algorithms providing a marginal
and/or inconsistent improvement in speed in a very narrow region (mainly
\SadaCL{} and \SadaIL) were left out. When \PDLBC{} and \PDLRP{} had very
similar performance, only one of them was chosen for the plot.

On \DNA, \Grammar{} was a good solution for small mutation rates, while \LZ{} was good with larger mutation rates. With more space available, \PDLBC{} became the fastest algorithm. \BruteD{} and \SadaID{} were often slightly faster than \PDL, when there was enough space available to store the document array. On \Concat{} and \Version{}, \PDL{} was usually a good mid-range solution, with \PDLRP{} being usually smaller than \PDLBC{}. The exceptions were the collections with $10$ base documents, where the number of variants ($1000$) was clearly larger than the block size ($256$). With no other structure in the collection, \PDL{} was unable to find a good grammar to compress the sets. At the large end of the size scale, algorithms using an explicit $\DA$ were usually the fastest choice.

\section{Conclusions}\label{section:conclusions}


Most document listing algorithms assume that the total number of occurrences of the pattern is large compared to the number of document occurrences. When documents are small, such as Wikipedia articles, this assumption generally does not hold. In such cases, brute-force algorithms usually beat dedicated document listing algorithms, such as Sadakane's algorithm and wavelet tree-based ones.

Several new algorithms have been proposed recently. \PDL{} is a fast and small
solution, effective on non-repetitive collections, and with
repetitive collections, if the collection is structured (e.g., incremental versions of base documents) or the average number of similar suffixes is not too large. Of the two \PDL{} variants, 
\PDLBC{} has a more stable performance, while \PDLRP{} is faster to build.
\Grammar{} is a small and moderately fast solution when the collection is repetitive but the individual documents are not. 
\LZ{} works well when repetition is moderate.

We adapted the \PDL{} structure for top-$k$ document retrieval. The new structure works well with 
repetitive collections, and is clearly 
the method of choice on the versioned \Revision{}.
When the collections are non-repetitive, brute-force algorithms remain competitive even on 
gigabyte-sized collections. While some dedicated algorithms can be faster, the price is
much higher space usage.





\bibliographystyle{plain}
\bibliography{paper}

\begin{thebibliography}{10}

\bibitem{CM13}
F.~Claude and I.~Munro.
\newblock Document listing on versioned documents.
\newblock In {\em Proc. SPIRE}, LNCS 8214, pages 72--83, 2013.

\bibitem{CN12}
F.~Claude and G.~Navarro.
\newblock Improved grammar-based compressed indexes.
\newblock In {\em Proc. SPIRE}, LNCS 7608, pages 180--192, 2012.

\bibitem{FN13}
H.~Ferrada and G.~Navarro.
\newblock A {L}empel-{Z}iv compressed structure for document listing.
\newblock In {\em Proc. SPIRE}, LNCS 8214, pages 116--128, 2013.

\bibitem{GKNPS13}
T.~Gagie, K.~Karhu, G.~Navarro, S.~J. Puglisi, and J.~Sir{\'e}n.
\newblock Document listing on repetitive collections.
\newblock In {\em Proc. CPM}, LNCS 7922, pages 107--119, 2013.

\bibitem{HNspire12.3}
C.~Hern{\'a}ndez and G.~Navarro.
\newblock Compressed representation of web and social networks via dense
  subgraphs.
\newblock In {\em Proc. SPIRE}, LNCS 7608, pages 264--276, 2012.

\bibitem{HSV09}
W.-K. Hon, R.~Shah, and J.~Vitter.
\newblock Space-efficient framework for top-$k$ string retrieval problems.
\newblock In {\em Proc. FOCS}, pages 713--722, 2009.

\bibitem{KN13}
R.~Konow and G.~Navarro.
\newblock Faster compact top-k document retrieval.
\newblock In {\em Proc. DCC}, pages 351--360, 2013.

\bibitem{LM00}
N.~J. Larsson and A.~Moffat.
\newblock Off-line dictionary-based compression.
\newblock {\em Proceedings of the IEEE Data Compression Conference},
  88(11):1722--1732, 2000.

\bibitem{Maekinen2010}
V.~M\"akinen, G.~Navarro, J.~Sir\'en, and N.~V\"alim\"aki.
\newblock Storage and retrieval of highly repetitive sequence collections.
\newblock {\em J. Comp. Bio.}, 17(3):281--308, 2010.

\bibitem{MM93}
U.~Manber and G.~Myers.
\newblock Suffix arrays: a new method for on-line string searches.
\newblock {\em SIAM J. Computing}, 22(5):935--948, 1993.

\bibitem{Mut02}
S.~Muthukrishnan.
\newblock Efficient algorithms for document retrieval problems.
\newblock In {\em Proc. SODA}, pages 657--666, 2002.

\bibitem{Naviwoca12}
G.~Navarro.
\newblock Indexing highly repetitive collections.
\newblock In {\em Proc. IWOCA}, LNCS 7643, pages 274--279, 2012.

\bibitem{NavACMcs14}
G.~Navarro.
\newblock Spaces, trees and colors: The algorithmic landscape of document
  retrieval on sequences.
\newblock {\em ACM Computing Surveys}, 46(4):article 52, 2014.

\bibitem{NM07}
G.~Navarro and V.~M{\"a}kinen.
\newblock Compressed full-text indexes.
\newblock {\em ACM Computing Surveys}, 39(1):art.~2, 2007.

\bibitem{NN12}
G.~Navarro and Y.~Nekrich.
\newblock Top-$k$ document retrieval in optimal time and linear space.
\newblock In {\em Proc. SODA}, pages 1066--1078, 2012.

\bibitem{NV12}
G.~Navarro and D.~Valenzuela.
\newblock Space-efficient top-k document retrieval.
\newblock In {\em Proc. SEA}, LNCS 7276, pages 307--319, 2012.

\bibitem{Sad07}
K.~Sadakane.
\newblock Succinct data structures for flexible text retrieval systems.
\newblock {\em J. Discrete Algorithms}, 5:12--22, 2007.

\bibitem{Wei73}
P.~Weiner.
\newblock Linear pattern matching algorithm.
\newblock In {\em Proc. 14th Annual IEEE Symposium on Switching and Automata
  Theory}, pages 1--11, 1973.

\end{thebibliography}


\clearpage
\appendix
\section*{Appendix}

\subsection*{Test Environment}

All implementations were written in C++ and compiled on g++ version 4.6.3. Our test environment was a machine with two 2.40 GHz quad-core Xeon E5620 processors (12~MB cache each) and 96~GB memory. Only one core was used for the queries. The operating system was Ubuntu 12.04 with Linux kernel 3.2.0.

\subsection*{Collections}

\begin{table}
\centering
\caption{Statistics for document collections. Collection size, CSA size without suffix array samples, number of documents, average document length, number of patterns, average number of occurrences and document occurrences, and the ratio of occurrences to document occurrences. For synthetic collections, most of the statistics vary greatly.}\label{table:collections}

\begin{tabular}{lrrcccccc}
\hline
\noalign{\smallskip}
Collection & \multicolumn{1}{c}{Size} & \multicolumn{1}{c}{CSA} & Documents & $n/d $ & Patterns & $\avg{occ}$ & $\avg{docc}$ & $occ/docc$ \\
\noalign{\smallskip}
\hline
\noalign{\smallskip}
\Page      &  110 MB &   2.58 MB &       60 & 1919382 &  7658 &   781 &     3 & 242.75 \\
           & 1037 MB &  17.45 MB &      280 & 3883145 & 20536 &  2889 &     7 & 429.04 \\
\noalign{\smallskip}
\Revision  &  110 MB &   2.59 MB &     8834 &   13005 &  7658 &   776 &   371 &   2.09 \\
           & 1035 MB &  17.55 MB &    65565 &   16552 & 20536 &  2876 &  1188 &   2.42 \\
\noalign{\smallskip}
\Enwiki    &  113 MB &  49.44 MB &     7000 &   16932 & 18935 &  1904 &   505 &   3.77 \\
           & 1034 MB & 482.16 MB &    90000 &   12050 & 19805 & 17092 &  4976 &   3.44 \\
\noalign{\smallskip}
\Influenza &  137 MB &   5.52 MB &   100000 &    1436 &  1000 & 24975 & 18547 &   1.35 \\
           &  321 MB &  10.53 MB &   227356 &    1480 &  1000 & 59997 & 44012 &   1.36 \\
\noalign{\smallskip}
\Swissprot &   54 MB &  25.19 MB &   143244 &     398 & 10000 &   160 &   121 &   1.33 \\
\noalign{\smallskip}
\DNA       &   95 MB &           &   100000 &         & 889--1000 \\
\Concat    &   95 MB &           & 10--1000 &         & 7538--15272 \\
\Version   &   95 MB &           &    10000 &         & 7537--15271 \\
\noalign{\smallskip}
\hline
\end{tabular}
\end{table}

\subsection*{Patterns}

\noindent {\bf Real collections.}
For \Page{} and \Revision{}, we downloaded a list of Finnish words from the Institute for the Languages in Finland, and chose all words of length $\ge 5$ that occur in the collection. 

For \Enwiki{}, we used search terms from an MSN query log with stop words filtered out. We generated $20000$ patterns according to term frequencies, and selected those that occur in the collection.

For \Influenza{}, we extracted $100000$ random substrings of length $7$, filtered out duplicates, and kept the $1000$ patterns with the largest $occ/docc$ ratios.

For \Swissprot{}, we extracted $200000$ random substrings of length $5$, filtered out duplicates, and kept the $10000$ patterns with the largest $occ/docc$ ratios.

\noindent {\bf Synthetic collections.}
For \DNA{}, patterns were generated with a similar process as for \Influenza{} and \Swissprot: take $100000$ substrings of length $7$, filter out duplicates, and choose the $1000$ with the largest $occ/docc$ ratios.

For \Concat{} and \Version{}, patterns were generated from the MSN query log in the same way as for \Enwiki. 


\end{document}